\documentclass[aps,prb,superscriptaddress,letterpaper,amsmath,amssymb,preprint,preprintnumbers,notitlepage]{revtex4}

\usepackage{amsmath}
\usepackage{amsthm}
\usepackage{amsxtra}
\usepackage{amstext}
\usepackage{amssymb}
\usepackage{bm}

\usepackage{graphicx}
\usepackage{IEEEtrantools}
\usepackage{latexsym}
\usepackage{lineno}

\makeatletter
\newcommand\ztag[1]{%
\def\@currentlabel{#1}%
\gdef\tmp{%
\addtocounter{equation}{-1}%
\def\theequation{#1}}%
\aftergroup\aftergroup\aftergroup\aftergroup\aftergroup\aftergroup
\aftergroup\aftergroup\aftergroup\aftergroup\aftergroup\aftergroup
\aftergroup\aftergroup\aftergroup\aftergroup\aftergroup\aftergroup
\aftergroup\aftergroup\aftergroup\aftergroup\aftergroup\aftergroup
\aftergroup\aftergroup\aftergroup\aftergroup\aftergroup\aftergroup
\aftergroup
\tmp}

\newcommand\vb[1]{\mathbf{#1}}

\theoremstyle{remark}

\makeatletter
\newcommand\RedeclareMathOperator{%
\@ifstar{\def\rmo@s{m}\rmo@redeclare}{\def\rmo@s{o}\rmo@redeclare}%
}
\newcommand\rmo@redeclare[2]{%
\begingroup \escapechar\m@ne\xdef\@gtempa{{\string#1}}\endgroup
\expandafter\@ifundefined\@gtempa
{\@latex@error{\noexpand#1undefined}\@ehc}%
\relax
\expandafter\rmo@declmathop\rmo@s{#1}{#2}}
\newcommand\rmo@declmathop[3]{%
\DeclareRobustCommand{#2}{\qopname\newmcodes@#1{#3}}%
}
\@onlypreamble\RedeclareMathOperator
\makeatother

\RedeclareMathOperator{\Re}{Re}
\RedeclareMathOperator{\Im}{Im}

\begin{document}
%\makeatother

%===========================
%   Title, Author, Affiliation
%===========================
\title{Gapless fluctuations and exceptional points in semiconductor lasers}
%\title{Gapless fluctuation regime of a semiconductor laser}
%\title{Fluctuation spectrum of laser with non-interacting electrons and holes in the $m=1$ channel}
\author{}
\date{\today}

%\date{}
% Activate to display a given date or no date (if empty), otherwise the current date is printed
%\date{May 22, 2022}

%\preprint{1}

\author{N.H.~\surname{Kwong}}
\affiliation{Wyant College of Optical Sciences, The University of Arizona, Tucson, AZ 85721}

\author{M.Em.~\surname{Spotnitz}}
\affiliation{Department of Physics, The University of Arizona, Tucson, AZ 85721}
\affiliation{Wyant College of Optical Sciences, The University of Arizona, Tucson, AZ 85721}

\author{R.~\surname{Binder}}
\affiliation{Wyant College of Optical Sciences, The University of Arizona, Tucson, AZ 85721}
\affiliation{Department of Physics, The University of Arizona, Tucson, AZ 85721}

%\pacs{???}

%======================================================
%   Abstract
%======================================================
\begin{abstract}
We analyze the spectrum of spatially uniform, single-particle fluctuation modes in the linear electromagnetic response of a semiconductor laser. We show that if the decay rate of the interband polarization, $\gamma_p$, and the relaxation rate of the occupation distribution, $\gamma_f$, are different, a gapless regime exists in which the order parameter $\Delta^{(0)} (k)$ (linear in the coherent photon field amplitude and the interband polarization) is finite but there is no gap in the real part of the single-particle fluctuation spectrum. The laser being a pumped-dissipative system, this regime may be considered a non-equilibrium analog of gapless superconductivity. We analyze the fluctuation spectrum in both the photon laser limit, where the interactions among the charged particles are ignored, and the more general model with interacting particles. In the photon laser model, the order parameter is reduced to a momentum-independent quantity, which we denote by $\Delta$. We find that, immediately above the lasing threshold, the real part of the fluctuation spectrum remains gapless when $0 < | \Delta | < \sqrt{2 / 27} \, | \gamma_f - \gamma_p |$ and becomes gapped when $| \Delta |$ exceeds the upper bound of this range. Viewed as a complex function of $|\Delta|$ and the electron-hole energy, the eigenvalue set displays some interesting exceptional point (EP) structure around the gapless-gapped transition. The transition point is a third-order EP, where three eigenvalues (and eigenvectors) coincide. Switching on the particle interactions in the full model modifies the spectrum of the photon laser model and, in particular, leads to a more elaborate EP structure. However, the overall spectral behavior of the continuous (non-collective) modes of the full model can be understood on the basis of the relevant results of the photon laser model.
\end{abstract}

\maketitle

\pagebreak

%\tableofcontents

%\narrowtext
%\begin{multicols}{2}

\section{Introduction}

A semiconductor quantum well (QW) microcavity laser is microscopically a highly coherent, driven-dissipative phase of electrons, holes, and photons.
 While a microcavity laser may operate similar to an edge-emitting laser where the Coulomb interaction between the charge carriers
 only affects properties like the lineshape of the gain spectra, but is not essential for the understanding of the fundamental lasing process
(e.g. Ref. [\onlinecite{chow-koch.99}]), there are other lasing regimes for which
 concepts constructed in theories of condensed phases of quantum fluids are commonly adapted to explain the basic laser properties.
One example is that of Bose-Einstein condensates (BEC) of exciton polaritons,
 \cite{%
imamoglu-etal.1996,%
moskalenko-snoke.00,%
eastham-littlewood.01,%
balili-etal.07,%
bajoni-etal.08,%
berney-etal.08,%
amo-etal.2009BEC,%
semkat-etal.09,%
deng-etal.10,%
menard-etal.14,%
schmutzler-etal.15,%
barachati-etal.2018,%
waldherr-etal.18,%
bao-etal.19,%
pieczarka-etal.2022prb,%
pieczarka-etal.2022,%
troue-etal.2023}
where exciton polaritons have properties similar to point bosons that can condense due to the nature of their bosonic quantum statistics.
Another example is that of
the so-called  polaritonic Bardeen-Cooper-Schrieffer (BCS) state
  \cite{%
comte-nozieres.82,%
keeling-etal.05,%
kremp-etal.08,%
kamide-ogawa.10,%
byrnes-etal.10,%
combescot-shiau.15,%
hu-liu.20,%
hu-etal.21},
 where the Coulomb interaction together with the light field in the cavity leads to bound electron-hole (e-h) pairs similar to Cooper pairs in superconductors, but unlike Cooper pairs there is a strong coupling between the Coulomb-correlated e-h pair and the cavity field.

An important concept in the original BCS theory \cite{bardeen-etal.1957} for superconductors in thermal equilibrium is that of an energy gap in the excitation spectrum of the superconducting state. The BCS gap, $\Delta$, which is related to the (attractive) interaction between electrons of opposite spin
and the resulting electron pairing
(Cooper pairs), can be used for an intuitive explanation of the absence of electrical resistivity.
In common textbook treatments
\cite{mattuck.1976,mahan.86,fetter-walecka.71,haken.73,madelung.78}, the Hamiltonian models used for the electrons lead to the result that the energy gap opens up at the normal-to-superconductor transition temperature, and the energy gap is proportional to the order parameter (anomalous Green's function). However, there are also classes of electron Hamiltonians \cite{maki.67} for which this `identification' of the energy gap with the order parameter is not valid. For example, in superconductors with dilute magnetic impurities
\cite{abrikosov-gorkov.61,reif-woolf.62,ambegaokar-griffin.65,maki.67,hansen.1968},
there is a gapless regime in which the order parameter is finite, and the system is superconducting, but the energy gap is zero.

In semiconductors interacting with strong electromagnetic fields that are tuned to interband transitions, an analogous gap is expected to open up in the single-particle spectrum.
\cite{galitskii-etal.70,elesin.71} 
We show in this paper that a gapless regime, analogous to that in superconductors, is present, quite generically, in a QW microcavity laser.

Since the laser is a non-equilibrium system, we need to define its spectral gap more precisely. For BCS superconductors at equilibrium, the gap is the minimum energy of single-particle excitations above the ground state. We consider a laser in a steady state, which is a driven-dissipative state, possibly far from equilibrium. As an analogy to the equilibrium case, we consider the fluctuation frequency spectrum of the steady state laser and possible gaps in the spectrum. The fluctuation matrix is obtained in a formulation of the laser's linear response to a weak probe (Section \ref{THz-linear-response.sec} and Appendix \ref{laser-fluctuation.sec}). Since the laser is dissipative, the fluctuation spectrum generally has a decay component, taking it off the real axis
in the plane of the complex energies of the fluctuation modes.
The gap we consider is the minimum finite frequency difference between two continuous fluctuation branches.

\begin{figure}[t]
\centering
\includegraphics[scale=0.5]{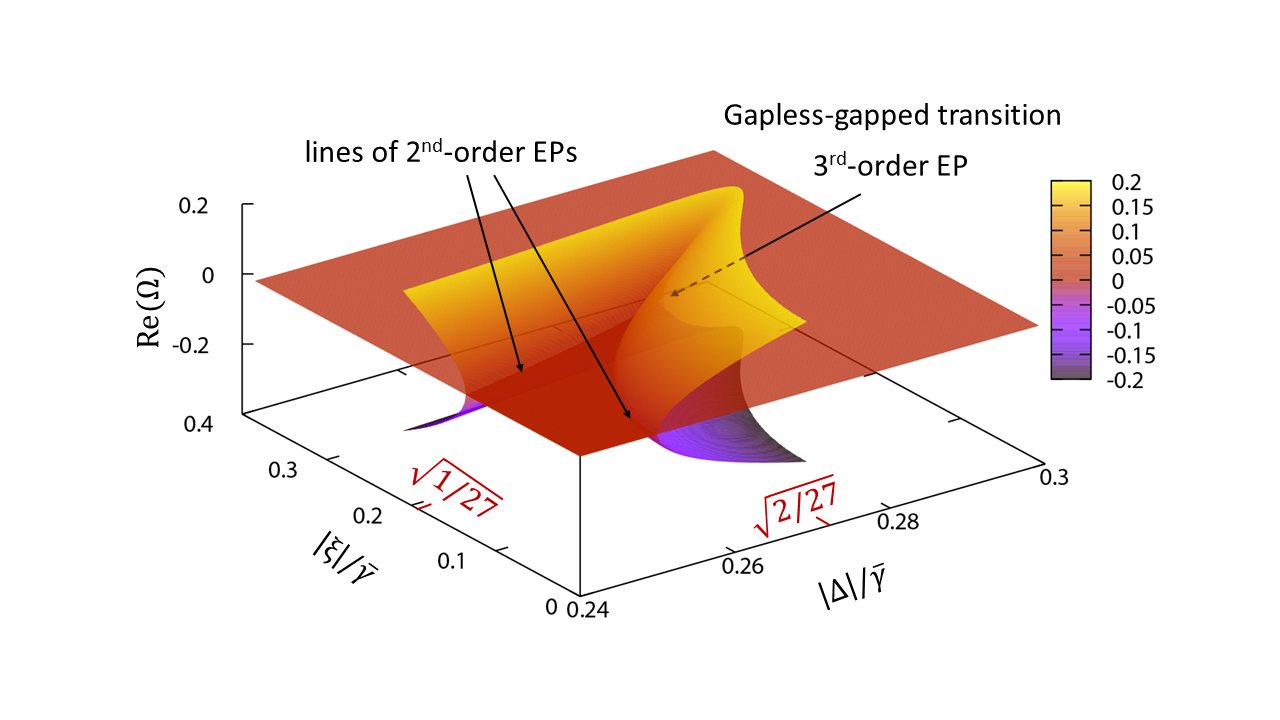}
\caption{(Color online)
3D plot of the real part of the fluctuation spectrum in the photon laser model (in which the interactions among charges are set to zero) as a function of the steady state order parameter $|\Delta|$ and the electron-hole energy measured from the laser frequency $\xi$. $\Omega = (\lambda + i \gamma_p) / \bar{\gamma}$, where $\lambda$ is the eigenvalue, $\bar{\gamma} = \gamma_f - \gamma_p$, $\gamma_p$ is the dephasing rate of the interband polarization, and $\gamma_f$ is the decay rate of the occupation fluctuation. All the energy variables are scaled to $\bar{\gamma}$. For each coordinate pair $( |\Delta| , \xi)$, there are three eigenvalues, which make up the three branches (surfaces) of the displayed spectrum. In the limited range of $|\Delta|$ plotted in the figure, two lines of second-order exceptional points (EP) meet at a third-order EP at $( |\Delta| / \bar{\gamma} , \xi / \bar{\gamma}) = ( \sqrt{2/27}  , \sqrt{1/27})$, which is the gapless-gapped transition point. $\bar{\gamma}$ is the critical parameter for the existence of a gapless regime: the range $0 < |\Delta| < \sqrt{2/27} \bar{\gamma}$ collapses to zero when $\bar{\gamma} \rightarrow 0$.
}
\label{3D-Real-Omega.fig}
\end{figure}

There has been an increased interest in exceptional points (EPs) recently which
are a common feature of non-Hermitian physics (see, e.g., Refs.
[\onlinecite{%
heiss.00,%
dembowski-etal.01,%
heiss.04,%
choi-etal.10,%
liertzer-etal.12,%
gao-etal.15,%
miri-alu.2019,%
pan-etal.19,%
sakhdari-etal.19,%
kawabata-etal.19,%
khurgin.2020,%
ohashi-etal.20,%
hanai-littlewood.20,%
ozturk-etal.2021,%
binder-kwong.2021,%
li-etal.2022,%
moiseyev.2011,%
ashida-etal.2020%
}]).
An EP is a point in the parameter space of a parametrized matrix where multiple eigenvalues and their eigenvectors coalesce, and the matrix is non-diagonalizable (defective).\cite{kato.95,heiss.00,ashida-etal.2020}
Some prior work focuses on optical systems similar to the model in this paper, but with some key differences.
Refs. [\onlinecite{joshi-galbraith.2018,am-shallem-etal.2015}], for example, discuss EPs in the Bloch equations that describe the time-evolution of a coherently-driven two-level system.
Our work differs in studying the Bloch equations for two energy bands (i.e., allowing for inhomogeneous broadening of the medium), instead of two discrete energy states. Equivalently, for degenerate two-level quantum systems, there is a single transition frequency, while in our case, we are studying a continuous range of laser-transition detunings.
The 
 physical difference between energy bands and generic broadening
 is more pronounced when we allow for Coulomb interactions among the energy-band states. 
 Nonetheless, we
 find that our key conclusions hold for both the interacting and noninteracting systems.
The cavity regime, i.e. where there is feedback on the photonic component from the matter component, has been investigated in Refs. [\onlinecite{hanai-etal.19,khurgin.2020,opala-etal.2023,rahmani-etal.2023}], and elsewhere. However, in this paper, we have 
 focused on
 the matter component, taking the semi-classical laser amplitude as an externally set constant. The spectra we study can be probed optically\cite{binder-kwong.2021} or in the terahertz (THz)\cite{spotnitz-etal.21}, by optically pumping a semiconductor sample outside a cavity. The spectra may also be sampled by a THz probe of a microcavity laser.\cite{spotnitz-etal.23} This latter thought experiment was the inspiration for this paper.

\begin{figure}[t]
\centering
\includegraphics[scale=0.5]{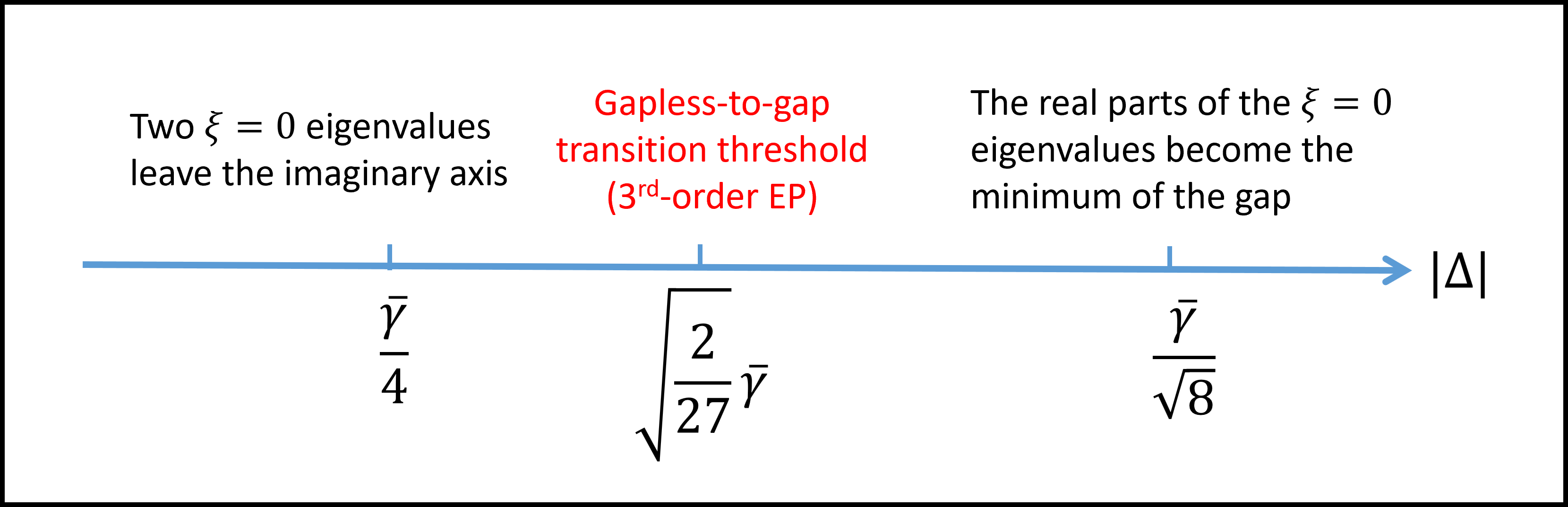}
\caption{(Color online)
Schematic drawing showing several threshold values of the order parameter $|\Delta|$ in the photon laser model. $\xi$ is the electron-hole energy measured from the laser frequency. $\xi = 0$ is the value at which the sum of the electron and hole band energies is in resonance with the laser frequency. See the caption of Fig. \ref{3D-Real-Omega.fig} for the definition of $\bar{\gamma}$. See Section \ref{free-eh.sec} for a detailed discussion.
}
\label{thresholds.fig}
\end{figure}

Our aim in this paper is to establish the condition for a gapless regime to occur and analyze the fluctuation spectrum, in particular the exceptional points structure, in detail.
Before presenting the theory and its analytical and numerical evaluation in the sections and appendices below, we believe it it useful to first give a relatively detailed summary of our results.

The unperturbed state to be probed is a lasing steady state, which we characterize by an energy order parameter $\Delta^{(0)} (\vb{k})= \Gamma_{e h} E^{(0)}_{\ell} + \Sigma_{\vb{k}^{\prime}} V_{\vb{k}-\vb{k}^{\prime}} p^{(0)}_{e h} (\vb{k}^{\prime})$. It is a combination of a photonic part (Rabi frequency) $\Gamma_{e h} E^{(0)}_{\ell}$, where $\Gamma_{e h}$ is the coupling that governs the transition from a cavity photon to an electron-hole pair and vice versa, and $E^{(0)}_{\ell}$ is the steady state laser field, and an electronic part (BCS 'gap function') $\Sigma_{\vb{k}^{\prime}} V_{\vb{k}-\vb{k}^{\prime}} p^{(0)}_{e h} (\vb{k}^{\prime})$, where $V_{\vb{k}-\vb{k}^{\prime}}$ is the electron-electron (electron-hole) Coulomb interaction, and $p^{(0)}_{e h} (\vb{k})$ is the interband polarization. $\vb{k}$ denoted the electron (and hole) momentum. In this paper, we refer to $\Delta^{(0)} (\vb{k})$, and also $E^{(0)}_{\ell}$ and $p^{(0)}_{e h} (\vb{k})$, as 'order parameters'. In our analysis, we first consider the photon laser limit
of the laser probed at THz frequency (e.g. \cite{ewers-etal.12,rice-etal.13,teich-etal.2014}, for THz response in the BEC limit see Ref. [\onlinecite{menard-etal.14}]).
The Coulomb interactions between the charged particles are ignored in this limit, and the electron-hole pairing is mediated by the coupling to the laser photon field. The THz probe drives intraband motions of the charges. The relative simplicity of this model allows analytic derivation of the most important results (Section \ref{free-eh.sec} and Appendix \ref{derivation.sec}). We find that a necessary condition for a gapless regime to occur is that the decay rate of the single-particle occupation fluctuations, $\gamma_f$, and that of the order parameter (interband polarization) fluctuations, $\gamma_p$, must be different. The difference of the two decay rates, $\bar{\gamma} = \gamma_f - \gamma_p$, is the critical parameter. (We assume in this paper that $\gamma_f \ge \gamma_p$ ($\bar{\gamma} \ge 0$). Similar conclusions hold for the case of $\bar{\gamma} < 0$.) With $V_{\vb{k}-\vb{k}^{\prime}}$ being set to zero, the order parameter $\Delta^{(0)} (\vb{k})$ is reduced to the Rabi frequency $\Delta^{(0)} (\vb{k}) \rightarrow \Delta = \Gamma E^{(0)}_{\ell}$, which is a $k$-independent constant. We follow the change in the fluctuation spectrum as the magnitude of $\Delta$ increases. The limit $\bar{\gamma}=0$ is the 'conventional' case, where there is no gapless regime: the threshold of lasing and the onset of the spectral gap coincide. More generally, when $\bar{\gamma} > 0$, the spectrum is gapless immediately above the lasing threshold. The onset of a gap occurs when $|\Delta|$ reaches a value of $\sqrt{\frac {2} {27}} \bar{\gamma}$. Around the transition, the spectrum shows some interesting behavior, which is illustrated in Fig. \ref{3D-Real-Omega.fig}, in which the real part of the spectrum is plotted as a function of $|\Delta|$ and the single-pair energy measured from the lasing frequency $\xi$. The plot shows that the spectrum has three branches (for each coordinate pair $( |\Delta| , \xi )$, there are three eigenvalues). Two lines of second-order EPs bound a region where all three eigenvalues have zero real parts (but their imaginary parts may be different). The two lines meet at a third-order EP at $( |\Delta| / \bar{\gamma} , \xi / \bar{\gamma} ) = (\sqrt{2/27} , \sqrt{1/27})$, which is the transition point between the gapless and the gapped regimes. There are thresholds in $|\Delta|$ other than the gapless-gapped transition which are of interest. These are shown in Fig. \ref{thresholds.fig}. The features in these two figures are explained and discussed in Section \ref{free-eh.sec} and Appendix \ref{derivation.sec}.

The analytical results obtained for the case without interaction also help us to identify similar phenomena of gapless lasing and the appearance of exceptional points, in particular a 3rd-order EP, in the case where electron-hole interaction is taken into account at the Hartree-Fock level, as discussed in Sec.
\ref{THz-full-model.sec}.
The spectra are complicated by the steady-state order parameter $\Delta^{(0)} (k)$ then being a function of $k$.
The results in the latter case are obtained from numerical solutions of the relevant equations of motion.
We note that the THz linear response of the laser was formulated and studied both at the photon laser limit in Ref. [\onlinecite{spotnitz-etal.21}] and with the more general model with interacting electrons and holes in Ref. [\onlinecite{spotnitz-etal.23}], but the issue of spectral gap opening is not discussed in those papers.

In Section \ref{THz-linear-response.sec}, we summarize our treatment of the THz linear response of the semiconductor laser. The fluctuation spectrum of the photon laser model (non-interacting electrons and holes) is analyzed through the gapless and gapped regimes in Section \ref{free-eh.sec}.
In Section \ref{THz-full-model.sec}, this
treatment is applied more generally to interacting electrons and holes.
 Appendix \ref{laser-fluctuation.sec} contains the equations of motion of the fluctuating fields and their reduction to the photon laser limit. Appendix \ref{derivation.sec} contains the algebraic detail of analytic results in the analysis of the photon laser fluctuation spectrum.

\section{Terahertz fluctuations of the quantum-well microcavity laser}
\label{THz-linear-response.sec}

We summarize here the dynamical model of the semiconductor microcavity laser on which we base our discussion. The model system is a quantum-well microcavity, in which electrons and holes in a single pair of conduction and valence bands in one or more quantum well(s) are coupled to a near-resonant optical field in the microcavity. The system is first prepared in a specified steady lasing state, and the fluctuations in this state are obtained as the linear response to a weak external probe. The band gap is set at the scale of $\sim 1 {\rm eV}$. The probe is set in the THz frequency range and so predominantly drives intraband motions rather than interband transitions. The optical conductivity of the THz probe, commonly chosen to characterize the linear response, is governed by the fluctuation eigenmodes of the laser. The eigenmodes and the optical conductivity were investigated numerically and their properties discussed in Refs. [\onlinecite{spotnitz-etal.21,spotnitz-etal.23}].

We treat the electron, hole, and photon dynamics at the level of the semiconductor Bloch equations (SBE). The photons are approximated as classical fields, and Coulomb correlations beyond the SBE, as well as interactions with the environment such as phonons are modeled phenomenologically as dephasing, relaxation and other gain/loss rates. The probe considered here is directed at normal incidence to the quantum well (QW) and so does not transmit any in-plane momentum to the fluctuations it creates. The dynamical fields, with this momentum restriction, are the electron distribution $f_e (\vb{k}, t) \equiv \langle \hat{a}^{\dagger}_{e \vb{k}} (t) \hat{a}_{e \vb{k}} (t) \rangle$, the hole distribution $f_h (\vb{k}, t) \equiv \langle \hat{a}^{\dagger}_{h \vb{k}} (t) \hat{a}_{h \vb{k}} (t) \rangle$, the interband polarization $p_{e h} (\vb{k}, t) \equiv \langle \hat{a}_{h (-\vb{k})} (t) \hat{a}_{e \vb{k}} (t) \rangle$, and the uniform cavity photon field $E_{\ell} (t)$. $\hat{a}_{e \vb{k}}$ is the annihilation operator of an electron in the conduction band orbital of momentum $\vb{k}$ and $\hat{a}_{h \vb{k}}$ annihilates a hole in the valence band.

The SBE-type equations governing the dynamical fields, and their expansion around a steady lasing state up to linear order in the probe, are discussed in Refs. [\onlinecite{spotnitz-etal.21,spotnitz-etal.23}]. For ease of reference, we include in Appendix \ref{laser-fluctuation.sec} the linear response equations and summarize the algebraic derivation in their reduction to an eigenvalue problem. For simplicity, we assume the electrons and the holes have the same physical attributes (effective mass, relaxation rates to quasi-equilibrium etc) so that $f_e (\vb{k}, t) = f_h (-\vb{k}, t)$ (if the two distributions are the same initially). We use $f_e (\vb{k}, t)$ to represent both distributions. We also assume that the laser steady state is set in a circularly polarized state and suppress the spin and polarization direction labels. Each field is written as a sum of two terms:
\begin{equation}
\Psi(\vb{k}, t) = \Psi^{(0)} (\vb{k}) + \Psi^{(1)} (\vb{k}, t) \quad , \quad \Psi = f_e , p_{e h} \quad , \quad E_{\ell} (t) = E^{(0)}_{\ell} + E^{(1)}_{\ell} (t)
\end{equation}
where the $(0)$ superscript denotes the lasing steady state and the $(1)$ superscript denotes the fluctuation response of the laser to the probe. As stated above, we expand the field equations up to first order in the probe. The unperturbed zeroth-order equations yield the steady-state fields which are used as input to the first-order fluctuation equations.
The fields $f^{(1)}_e (\vb{k}, t)$ and $p^{(1)}_{e h} (\vb{k}, t)$ are expanded in an orbital angular momentum basis with the quantization axis being normal to the QW's plane: $\Psi^{(1)} (\vb{k} , t) = \sum_{m \in \mathbb{Z}} \Psi^{(1)} (k,m,t) e^{im\theta_k}$, $\vb{k} = (k , \theta_k)$, $\Psi^{(1)} = f^{(1)}_e , p^{(1)}_{e h}$. Since the electric field fluctuation $E^{(1)}_{\ell} (t)$ is independent of $\vb{k}$, its $m=0$ component is the only non-zero term in such an expansion. The normally-incident probe is assumed to be a plane wave. If it is tuned to the optical frequency range and so drives interband transitions, only the $m=0$ components, $(f^{(1)}_e (k,0,t), p^{(1)}_{e h} (k,0,t), E^{(1)}_{\ell} (t))$, participate in the response \cite{binder-kwong.2021}. For THz-frequency probes considered in this paper, which mainly drives intraband electronic motion, it is shown in Ref. [\onlinecite{spotnitz-etal.23}] that only the $m = \pm 1$ angular components, $(f^{(1)}_e (k,\pm 1,t), p^{(1)}_{e h} (k,\pm 1,t)$ are triggered. Compared to the case of optical-frequency probes, the absence of the photon component in the THz fluctuations is a substantial simplification technically, allowing a much more thorough analysis. The insight obtained here is also helpful for the interpretation of the fluctuation spectra obtained numerically for optical-frequency probes \cite{binder-kwong.2021}. In the next two sections, we analyze the THz fluctuation spectra in the photon laser model, where the interactions among the charges are tuned off, and the continuum part(s) of the spectra in the fully interacting model.

\begin{figure}[t]
\centering
\includegraphics[scale=0.5]{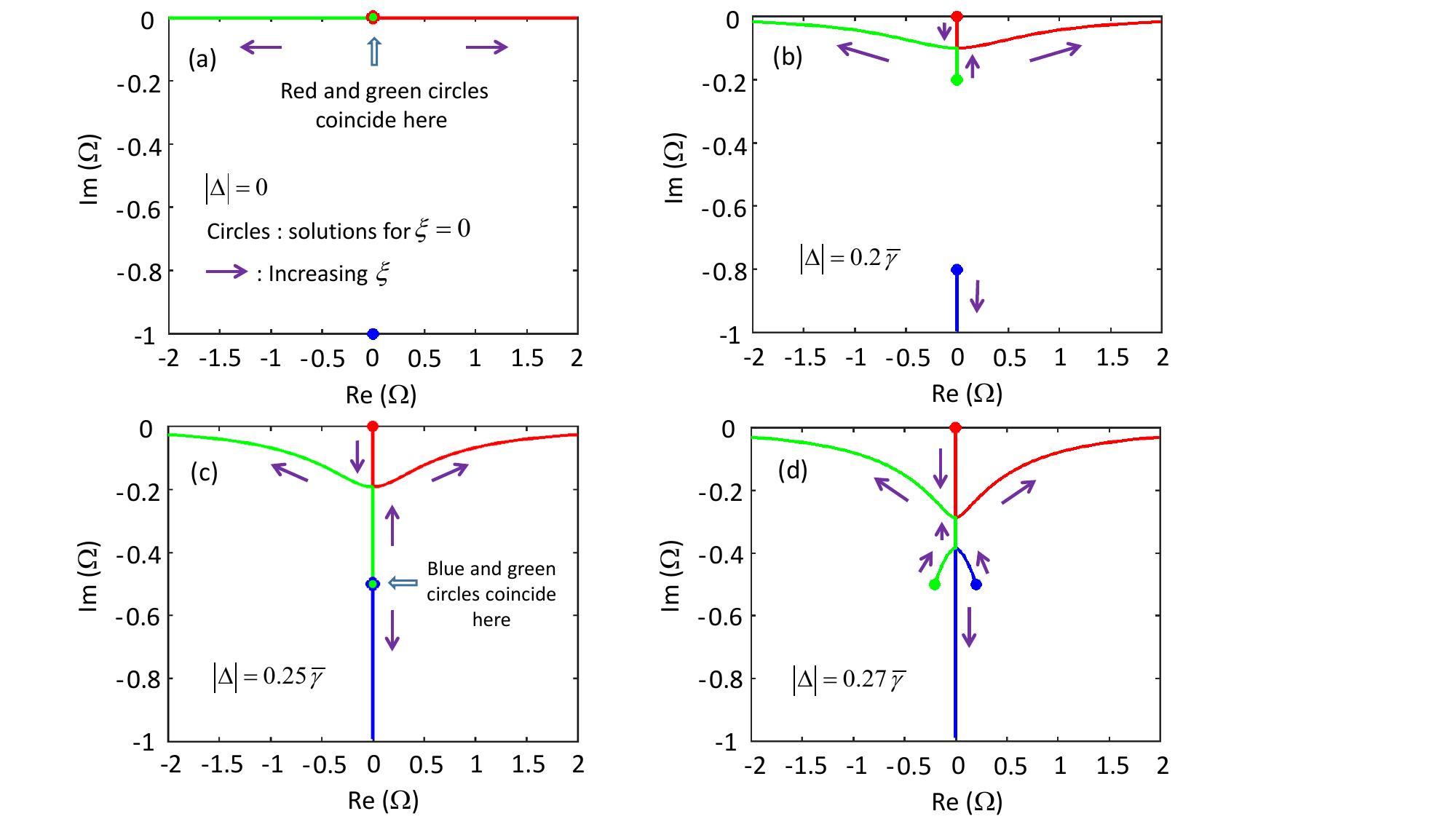}
\caption{(Color online) Fluctuation eigenvalue spectra in the gapless regime for the non-interacting electron-hole model. The eigenvalue $\bar{\lambda} = \lambda + i \gamma_p$ is scaled to the decay rate difference $\bar{\gamma} = \gamma_f - \gamma_p$: $\Omega = \bar{\lambda} / \bar{\gamma}$. Each panel shows the spectrum calculated with the value of the order parameter $| \Delta |$ given in the panel. The spectrum consists of three continuous branches color-coded as red, blue, and green. Each branch traces an eigenvalue when $\xi$, the electron-hole energy measured from the lasing frequency, changes. See Eq. (\ref{M-def.equ}) for the definitions of $\Delta$ and $\xi$. The solid circles mark the eigenvalues at $\xi=0$, and the arrows indicate the directions of increasing $\xi$. The spectrum contains one second-order exceptional point (EP) when $|\Delta| / \bar{\gamma} < 1 / 4$ (panels (b) and (c)), and a second EP emerges when $|\Delta| / \bar{\gamma}$ exceeds $1 / 4$ (panel (d)). }
	\label{spectra-1.fig}
\end{figure}

\begin{figure}[t]
\centering
\includegraphics[scale=0.5]{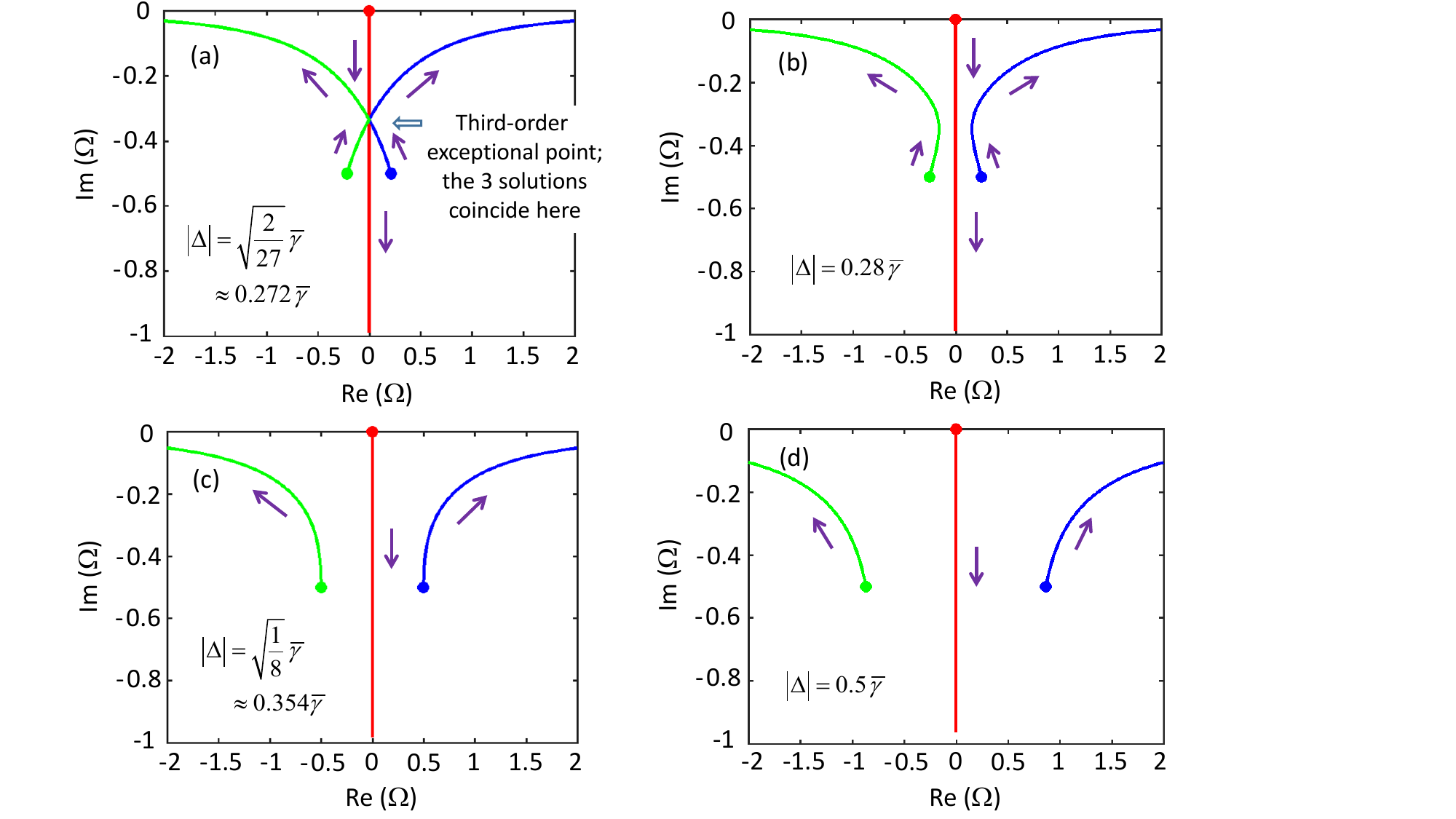}
%includegraphics[scale=0.5,bb = 0 0 900 500]{Fig-04-spectra-free-2.pdf}
\caption{Transition from the gapless regime to the gapped regime (see Fig. 1 for the meanings of symbols). Panel (a) shows the transition at $|\Delta| / \bar{\gamma} = \sqrt{2 / 27}$. At this point, the two second-order EPs shown in Fig. \ref{spectra-1.fig}(d) coincide to form a third-order EP. At higher values of $|\Delta|$ (panels (b), (c), and (d)), a gap appears in ${\rm Re} \Omega$. When $|\Delta| / \bar{\gamma} \ge \sqrt{1 / 8}$ (panels (c) and (d)), the gap equals the value of ${\rm Re} \Omega$ at $\xi = 0$, where the free electron-hole energy is at resonance with the lasing frequency. }
\label{spectra-2.fig}
\end{figure}

\section{THz fluctuation spectrum in the photon laser limit: gapless regime and exceptional points structure}
\label{free-eh.sec}

In this section, we analyze the THz fluctuation spectrum in the photon laser limit, where the Coulomb interaction among the charge carriers are ignored. The algebraic steps in reducing the response equations to this limit are summarized in Appendix \ref{laser-fluctuation.sec}. As stated above, the angular momentum channels are block-decoupled, and only the $m = \pm 1$ channels are accessed by a (plane-wave) THz probe. The response equations for different $k = | \vb{k} |$ values are also block-decoupled. The equation for each $k$, in a three-dimensional column vector form, are (Appendix \ref{laser-fluctuation.sec})
\begin{equation}
i \hbar \frac {\partial} {\partial t} \vec{x} (k,1,t) = M (k) \vec{x} (k,1,t) + \vec{S} (k,1,t)
\end{equation}
where
\begin{equation}
\vec{x} (k,1,t) =
\begin{pmatrix}
 p^{(1)}_{e h} (k,1,t) \\ p^{(1) \ast}_{e h} (k,-1,t) \\ f^{(1)}_{e} (k,1,t)
\end{pmatrix}
\quad , \quad
\vec{S} (k,1,t) =
\begin{pmatrix}
S (k,1,t) \\ - S^{\ast} (k,-1,t) \\ 0
\end{pmatrix}
\end{equation}
and the matrix $M$ is
\begin{equation}
\label{M-def.equ}
M =
\begin{pmatrix}
\xi(k) - i \gamma_p & 0 & 2 \Delta \\
0 & - \xi(k) - i \gamma_p & -2 \Delta^{\ast} \\
\Delta^{\ast} & - \Delta & - i \gamma_f
\end{pmatrix}
\quad , \quad
\xi(k) = \frac {\hbar^2 k^2} {m_e} + E_g - \hbar \omega_{\ell} \quad , \quad \Delta = \Gamma_{e h} E^{(0)}_{\ell}
\end{equation}
The $m = \pm 1$ components of the photon field vanish, $E^{(1)}_{\ell} (k,\pm 1,t) = 0$, leaving only the electron and hole fields in the response.
In the matrix $M$, $m_e$ is the common effective mass of the conduction band and the valence band, $E_g$ is the band gap, $\omega_{\ell}$ is the lasing frequency, $\gamma_p$ is the dephasing rate of the polarization $p_{e h}$, and $\gamma_f$ is the relaxation/loss rate of the density distribution $f_e$. $\Gamma_{e h}$ is the coupling that governs the transition from a cavity photon to an electron-hole pair and vice versa, $E^{(0)}_{\ell}$ is the steady state laser field, and their product gives the steady state order parameter $\Delta$. $\xi (k)$ is the electron-hole energy measured from the laser frequency. $S(k,1,t)$ denotes the THz probe field.

In the following, we discuss the behavior of the eigenmode spectrum of $M$ as a function of the parameters.
The eigenvalue is denoted generically by $\lambda$.
The eigenvalues satisfy the equation
\begin{equation}\label{eigen.equ}
{\rm det} \left[ \lambda - M \right] = ( \bar{\lambda}^2 - \xi^2 )( \bar{\lambda} + i \bar{\gamma} )
- 4 |\Delta|^2 \bar{\lambda} = 0
\end{equation}
where we have used the notation
\begin{equation}
\bar{\lambda} = \lambda + i \gamma_p \quad , \quad \bar{\gamma} = \gamma_f - \gamma_p
\end{equation}
We treat the loss rate difference $\bar{\gamma}$, the magnitude of the order parameter $|\Delta|$ and the electron-hole energy $\xi$ as parameters to be varied. The other parameters, $\hbar \omega_{\ell}$, $E_g$, and $\gamma_p$, are fixed. The steady state is prepared to have the lasing frequency larger than the band gap, $\hbar \omega_{\ell} - E_g > 0$.
Eq. (\ref{eigen.equ}) is a complex cubic equation, and explicit (though rather cumbersome) solutions are available (e.g. [\onlinecite{abramowitz-stegun.72}]). We show some computed spectra in Figs. \ref{spectra-1.fig} and \ref{spectra-2.fig}. In each panel in the figures, three sets of eigenvalues $\bar{\lambda}$ are plotted in the complex plane for a fixed value of the ratio $|\Delta| / \bar{\gamma}$, $\bar{\gamma}$ being assumed to be positive. Each set is a continuous curve of an eigenvalue parameterized by $\xi$ (or equivalently $k$). (The $\lambda$ spectrum is obtained by shifting the $\bar{\lambda}$ spectrum down by $- i \gamma_p$.)
Some analytic results on the properties of the spectrum are also derived to organize our understanding of the displayed features in the figures. These analytic results are used here, and the algebraic details of their derivation are collected in Appendix \ref{derivation.sec}.

\subsubsection{The special case of equal decay rates: $\bar{\gamma} = \gamma_f - \gamma_p = 0$}
\label{equal-rate.sec}

When the decay rates of the single-particle occupation fluctuation and the interband polarization fluctuation are set to be equal: $\gamma_f = \gamma_p \Rightarrow \bar{\gamma} = 0$, the spectrum is simply
\begin{equation}
\bar{\lambda} ({\bar{\gamma}=0}) = 0, \pm \sqrt{\xi^2 + 4 | \Delta |^2}
\end{equation}
$\bar{\lambda}$ is real for all $\xi$. The gap in the spectrum, denoted by $E^{gap}_{eh}$, is given by the minimum magnitude of the non-zero branches of $\bar{\lambda}$, which is the magnitude of the eigenvalue at $\xi = 0$:
\begin{equation}
E^{gap}_{eh} = 2 |\Delta |
\end{equation}
So there is no gapless regime, where the system is lasing ($|\Delta|$ is finite) but $E^{gap}_{eh} = 0$, in this case.

\subsubsection{Overall distribution of the eigenvalues}

It can be readily verified that if $\bar{\lambda}$ is a solution to the eigenvalue equation (\ref{eigen.equ}), then so is $- \bar{\lambda^{\ast}}$ (See Ref. [\onlinecite{spotnitz-etal.23}], Appendix A.). This implies that, for each $k$, either all three eigenvalues $\lambda$ are imaginary or one eigenvalue
is imaginary and the other two have finite real parts and are symmetrically placed, one on each side of the imaginary axis. This symmetry about the imaginary axis also applies to the eigenvalue set of the more general case with interacting electrons and holes \cite{spotnitz-etal.23}.

When $\bar{\gamma} > 0$, $\bar{\lambda} (k)$ on the imaginary axis are confined to the interval $[ - i \bar{\gamma} , 0 ]$ for all $k$. For $\bar{\lambda}(k)$ off the imaginary axis,
${\rm Im \bar{\lambda} (k)} \in [ - \bar{\gamma} / 2 , 0 ]$ for all $k$. These bounds are derived in Appendix \ref{derivation.sec} (Property (1)). We note that these bounds collapse the area containing the eigenvalues to a horizontal line when $\bar{\gamma}$ goes to zero.

\subsubsection{The gapless regime for $\bar{\gamma} > 0$}

Immediately above the lasing threshold, the order parameter $|\Delta|$ increases, starting from zero, with the pump intensity. The fluctuation spectrum is gapless within the range
\begin{equation}
0 < \frac {|\Delta|} {\bar{\gamma}} < \sqrt{\frac {2} {27}}
\end{equation}
Figs. \ref{spectra-1.fig} and \ref{spectra-2.fig}(a) show in detail the changes in the spectra as $|\Delta|$ increases within this regime. The solid circles in each panel are the three eigenvalues at a $k$ value, denoted by $k_{\ell}$, such that $\xi (k_{\ell}) = \hbar^2 k_{\ell}^2 / m_e + E_g - \hbar \omega_{\ell} = 0$ where the electron-hole band energy is at resonance with the laser frequency. Since Eq. (\ref{eigen.equ}) depends on $\xi^2$, the parametrization of the eigenvalues goes from $\xi = 0$ upwards, and so the solid circles are the starting points of the three eigenvalue curves on the plots. The general behavior of the eigenvalues at the limits $\xi = 0$ and $\xi \rightarrow \infty$ are derived in Appendix \ref{derivation.sec} (Properties (2) and (3)).
When $| \Delta | < \bar{\gamma} / 4$, all three solid circles stay on the imaginary axis (Fig. \ref{spectra-1.fig}(b)):
\begin{equation}
\bar{\lambda} (\xi=0) = 0,  - i \left[ \frac {\bar{\gamma}} {2} \pm \sqrt{\left( \frac {\bar{\gamma}} {2}
\right)^2 - 4 | \Delta |^2} \right] \quad {\rm for} \quad | \Delta | < \frac {\bar{\gamma}} {4}
\end{equation}
As $| \Delta |$ increases, the last two solutions move towards each other and coincide at the value of $- i \bar{\gamma} / 2$ when $| \Delta | = \bar{\gamma} / 4$ (Fig. \ref{spectra-1.fig}(c)).
When $| \Delta | > \bar{\gamma} / 4$, the two solid circles move out in opposite directions along the line ${\rm Im \bar{\gamma}} = - \bar{\gamma} / 2$:
\begin{equation}
\bar{\lambda} (\xi=0) = 0, \pm \sqrt{ 4 | \Delta |^2 - \left( \frac {\bar{\gamma}} {2}
\right)^2 } - i \frac {\bar{\gamma}} {2}  \quad {\rm for} \quad | \Delta | > \frac {\bar{\gamma}} {4}
\end{equation}
Before $|\Delta|$ reaches $\sqrt{2 / 27} \bar{\gamma} \approx 0.272 \bar{\gamma}$, there is still no gap in ${\rm Re} \bar{\lambda}$ even though two eigenvalues at $\xi=0$ leave the imaginary axis. Fig. \ref{spectra-1.fig}(d) shows the spectrum at $|\Delta| = 0.27 \bar{\gamma}$ as an example. Starting from the two solid circles with finite real parts, the two eigenvalue branches move towards and meet on the imaginary axis as $\xi$ increases. Beyond that point, one eigenvalue moves up while the other moves down the imaginary axis. The upward-going eigenvalue then meets with the third eigenvalue and the two move off the imaginary axis. If we view the matrix $M$ as a complex matrix function of $\xi$ and $\Delta$, then the two points where two eigenvalues meet are second-order exceptional points (EP) \cite{heiss.04,miri-alu.2019,hanai-etal.19}.

As $| \Delta |$ increases, the two second-order EPs in the spectrum shift toward each other and coincide at $| \Delta | = \sqrt{2 / 27} \, \bar{\gamma}$. This value of $|\Delta|$ also marks the threshold for transition from the gapless regime to the gapped regime. The spectrum at this threshold is shown in Fig. \ref{spectra-2.fig}(a). It is verified in Appendix \ref{derivation.sec} (Property (4)) that the three eigenvalue branches coincide at the parametric point $\xi =  \frac {\bar{\gamma}} {\sqrt{27}}$ with the value $\bar{\lambda} = - i \frac {\bar{\gamma}} {3}$.
This point where the three eigenvalues meet is a third-order exceptional point. To see this, we check that the eigenvectors of the three branches are also the same, and the common eigenvector is self-orthogonal (the inner product of the left and right eigenvectors equals zero). The right and left eigenvectors at the EP are denoted by $\vec{\phi}_R$ and $\vec{\phi}_L$ respectively (the left eigenvector satisfies $\vec{\phi}^{\, \dag}_{L} (\lambda - M) = 0$). They are given by (without normalization)
\begin{equation}
\vec{\phi}_R =
\begin{pmatrix}
1 \\
e^{i 2 (\pi / 3 - \theta)} \\
- \frac {1} {\sqrt{2}} e^{i (\pi / 3 - \theta)}
\end{pmatrix}
\quad , \quad
\vec{\phi}_L =
\begin{pmatrix}
1 \\
e^{- i 2 (\pi / 3 + \theta)} \\
- \sqrt{2} e^{- i (\pi / 3 + \theta)}
\end{pmatrix}
\end{equation}
where $\theta$ is the phase of $\Delta$: $\Delta = | \Delta | e^{i \theta}$. It can be checked that $\vec{\phi}^{\, \dag}_L \vec{\phi}_R =0$.
The exceptional points structure of the spectrum near the gapless-gapped transition point is shown in a 3D plot in Fig. \ref{3D-Real-Omega.fig}.

%There are also second-order EP's, where two eigenvalues coincide, in the gapless regime. One EP is present when $|\Delta| < \frac {\bar{\gamma}} {4}$, as shown in Fig. \ref{spectra-1.fig}(b), and there are two EP's within the range $\frac {\bar{\gamma}} {4} < |\Delta| < \sqrt{\frac {2} {27}} {\bar{\gamma}}$, as shown in \ref{spectra-1.fig}(d).

\subsubsection{Gapped regime}
A gap appears in ${\rm Re} \bar{\lambda}$ when $| \Delta |$ exceeds $\sqrt{2 / 27} \, \bar{\gamma}$. As shown in Fig. \ref{spectra-2.fig}(b), two branches separate from the imaginary axis. The minimum value of $| {\rm Re} \bar{\lambda} |$, which gives the value of the gap, is obtained at a non-zero $\xi$. As $|\Delta|$ increases further, this minimum value gains on $| {\rm Re} \bar{\lambda} (\xi=0)|$, and from $| \Delta | = \frac {\bar{\gamma}} {\sqrt{8}}$ on, the resonance points $\xi=0$ (solid circles) give the value of the gap (panels (c) and (d) of Fig. \ref{spectra-2.fig}), which is
\begin{align}
E^{gap}_{eh} &= \sqrt{4 | \Delta |^2 - \left( \frac {\bar{\gamma}} {2} \right)^2} \quad {\rm for} \quad
| \Delta | > \frac {\bar{\gamma}} {\sqrt{8}} \\
& \approx 2 | \Delta | \quad {\rm when} \quad | \Delta| \gg \frac {\bar{\gamma}} {4}
\label{large-Delta.equ}
\end{align}
Eq. (\ref{large-Delta.equ}) shows that the result of the case of $\bar{\gamma}=0$ (Section \ref{equal-rate.sec}) is recovered in the limit of large $\frac {|\Delta|} {\bar{\gamma}}$, as would be expected. That $| \Delta | = \frac {\bar{\gamma}} {\sqrt{8}}$ is the transition value is proved in Appendix \ref{derivation.sec} (Property (5)).

\subsubsection{Discussion}

As shown above, the essential condition for the the presence of a gapless regime is that the fluctuations in the order parameter $p^{(1)}_{e h}$ and those in the single-particle occupation $f^{(1)}_e$ decay at different rates: $\gamma_f \neq \gamma_p$. (We show this only for $\gamma_f > \gamma_p$, but it is also valid for $\gamma_f < \gamma_p$.) This condition bears some similarity to the condition for gapless superconductivity in alloys with magnetic impurities \cite{abrikosov-gorkov.61,ambegaokar-griffin.65,maki.67}, where the spin-orbit term in the Hamiltonian leads to a gapless regime by causing a difference in the relaxation rates in the order parameter and the single-particle propagator. Another possible point for comparison is time-reversal invariance breaking. In the case of the superconductor with magnetic impurities, the fact that the spin-orbit term breaks time-reversal invariance is considered an important factor for gaplessness \cite{maki.67}. Time reversal invariance is obviously broken in our laser system. However, the nature of the invariance breaking is different in the two cases. In the superconductor case, one considers the ground state or a thermal equilibrium state. The Hamiltonian containing the spin-orbit term is Hermitian, and the Schr\"{o}dinger equation containing it does not have a preferred time direction. In contrast, the equations of motion in our laser case distinguish between going forward and backward in time. The exceptional points in our spectra are present because of the non-Hermitian nature of the laser system.

\begin{figure}[ht]
\centering
\includegraphics[scale=0.45,trim= 15cm 1cm 15cm 2cm]{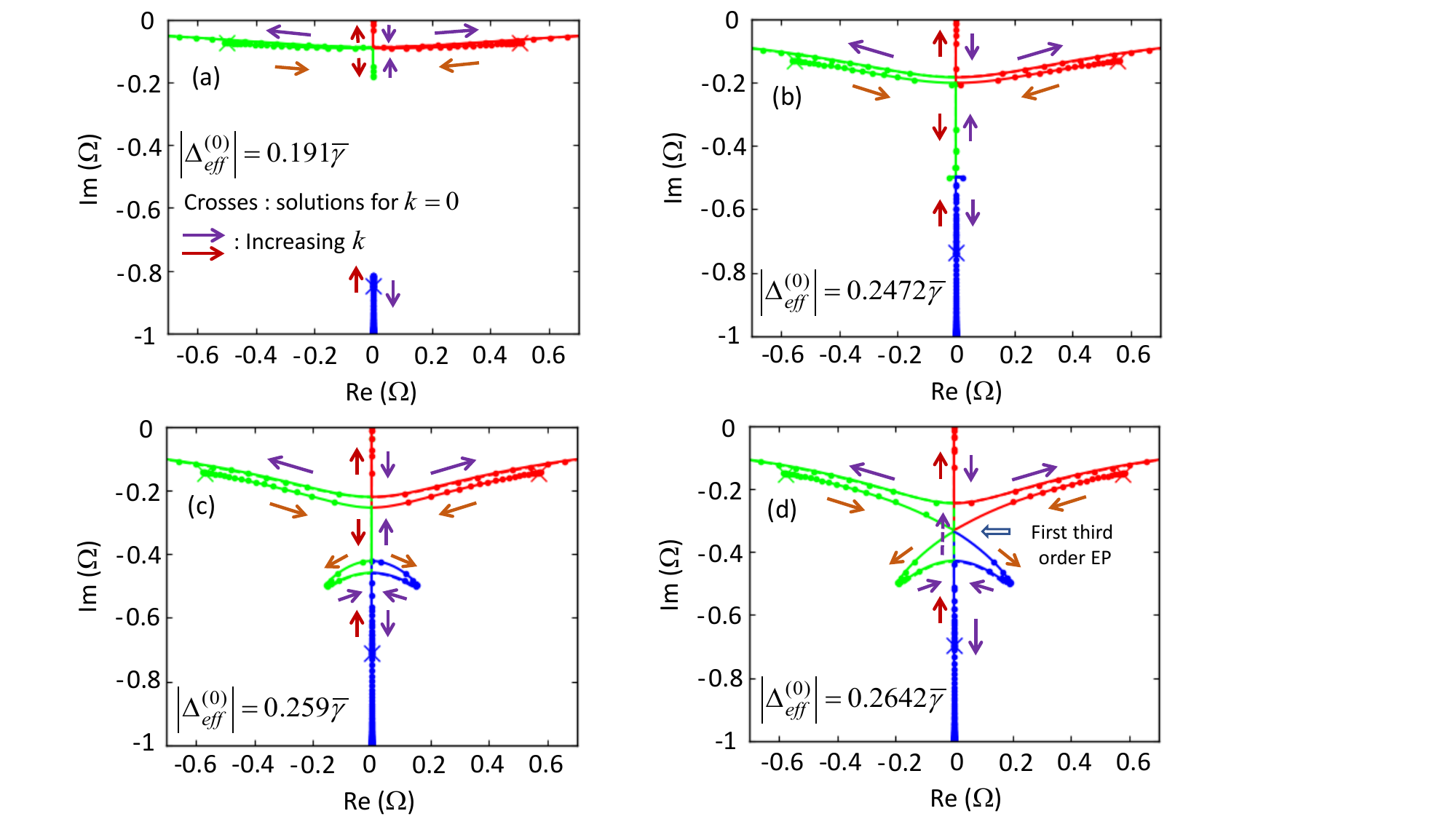} % trim doesn't seem to work
%includegraphics[scale=0.43,bb = 0 0 900 500]{Fig-05-spectra-Coulomb-3.pdf}
%                                      left bottom  right  top
%   after -90 rotation:  top  left  bottom right
\caption{(Color online) Fluctuation eigenvalue spectra in the gapless regime for the electron-hole model with screened Coulomb interactions. The eigenvalue $\bar{\lambda} = \lambda + i \gamma_p$ is scaled to the decay rate difference $\bar{\gamma} = \gamma_f - \gamma_p$: $\Omega = \bar{\lambda} / \bar{\gamma}$. In each panel, the dots are eigenvalues calculated by numerical diagonalization on a discretized $k$ grid. The solid curves are calculated neglecting the $k$-mixing terms in the matrix. The non-interacting electron-hole model with an effective constant order parameter $| \Delta^{(0)}_{\textit{eff}} |$ is used to fit the eigenvalue curves.
The best-fit value of $| \Delta^{(0)}_{\textit{eff}} |$ is given in each panel as a measure of the overall strength of the $k$-dependent order parameter $| \tilde{\Delta}^{(0)} (k)|$. The functions $| \tilde{\Delta}^{(0)} (k)|$ for the cases in this figure and Fig. \ref{spectra-4.fig} are plotted in Fig. \ref{Delta-vs-k.fig}. The best-fit eigenvalues of the non-interacting model are plotted as dashed lines. The crosses mark the three eigenvalues at $k=0$, and the arrows point in the direction of motion of the eigenvalues when $k$ increases. The brown arrows are for $k < k_{\ell}$ and the purple arrows are for $k > k_{\ell}$, where $k_{\ell}$ is the value at which $\tilde{\xi} ( k_{\ell} ) = 0$. There are two second-order exceptional points (EP) in panel (b) and four EPs in panel (c). In panel (d), the EP structure consists of a third order EP between two second order EPs. The dashed purple arrow indicates the flow of the green branch between the two second order EPs. The third order EP results from the merging of the two inner EPs in panel (c). (See text for further details.) }
\label{spectra-3.fig}
\end{figure}

\begin{figure}[ht]
\centering
\includegraphics[scale=0.45,trim= 10cm 0cm 10cm 0cm]{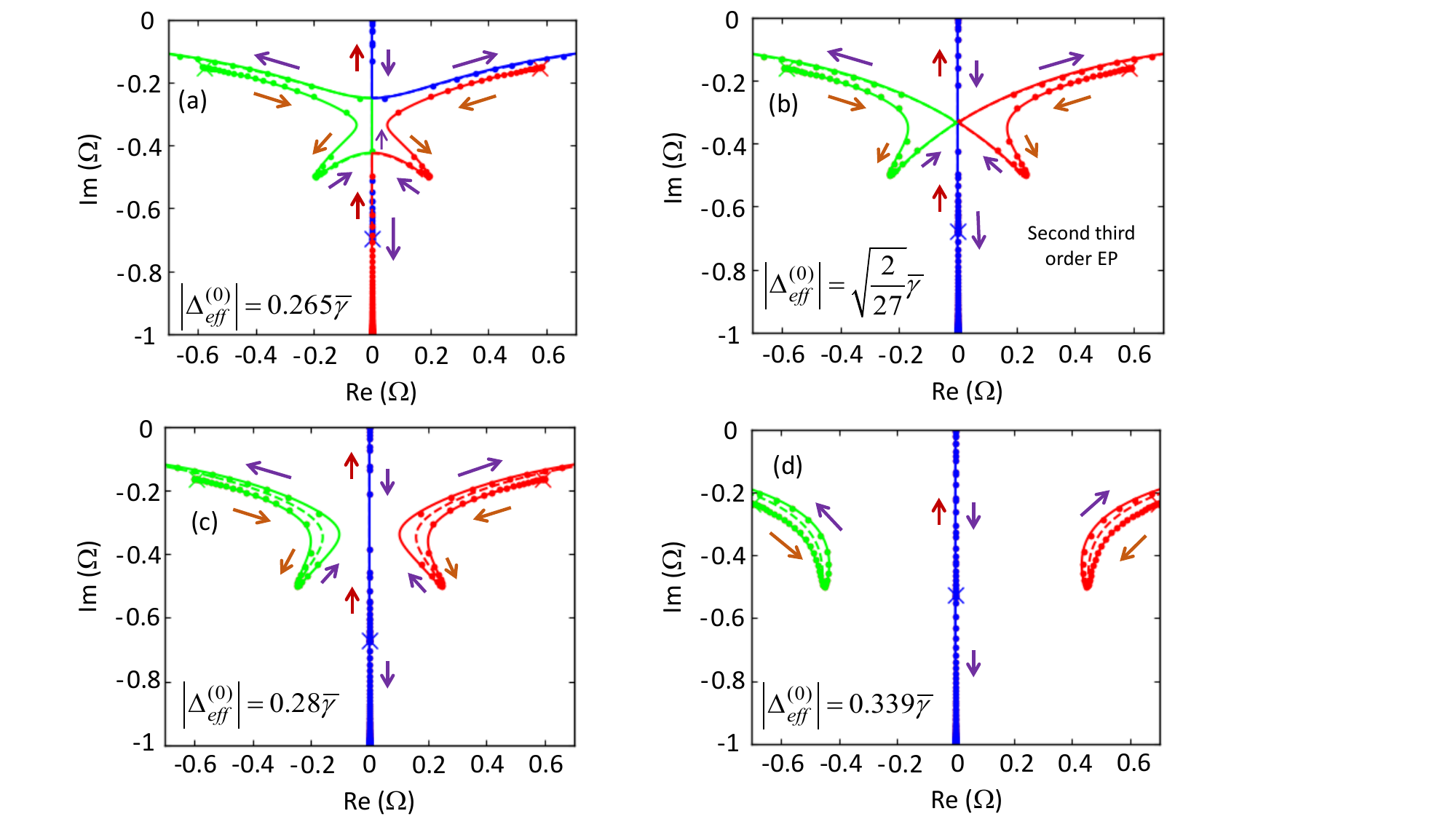}
%                                      left bottom  right  top
%   after -90 rotation:  top  left  bottom right
%\includegraphics[scale=0.5,bb = 0 0 900 500]{Fig-06-spectra-Coulomb-4.pdf}
\caption{Transition from the gapless regime to the gapped regime (see Fig. \ref{spectra-3.fig} for the meanings of the dots, the curves, the arrows, and $| \Delta^{(0)}_{\textit{eff}} |$). Partly visible here, but covered entirely by the solid lines in Fig. \ref{spectra-3.fig}, are dashed lines, which are the noninteracting eigenvalue spectra calculated using $\left| \Delta^{(0)}_{\textit{eff}} \right|$. Panel (a) shows that, when $| \Delta^{(0)}_{\textit{eff}} |$ exceeds the value in Fig. \ref{spectra-3.fig}(d), two branches separate from the imaginary axis, leaving two second order EPs in the spectrum. These two EPs merge in panel (b) to form a second third order EP, which marks the gapless-gapped transition. Beyond this point, two gapped branches of the spectrum appear (panels (c) and (d)).
}
\label{spectra-4.fig}
\end{figure}

\begin{figure}[ht]
\centering
\includegraphics[scale=0.5]{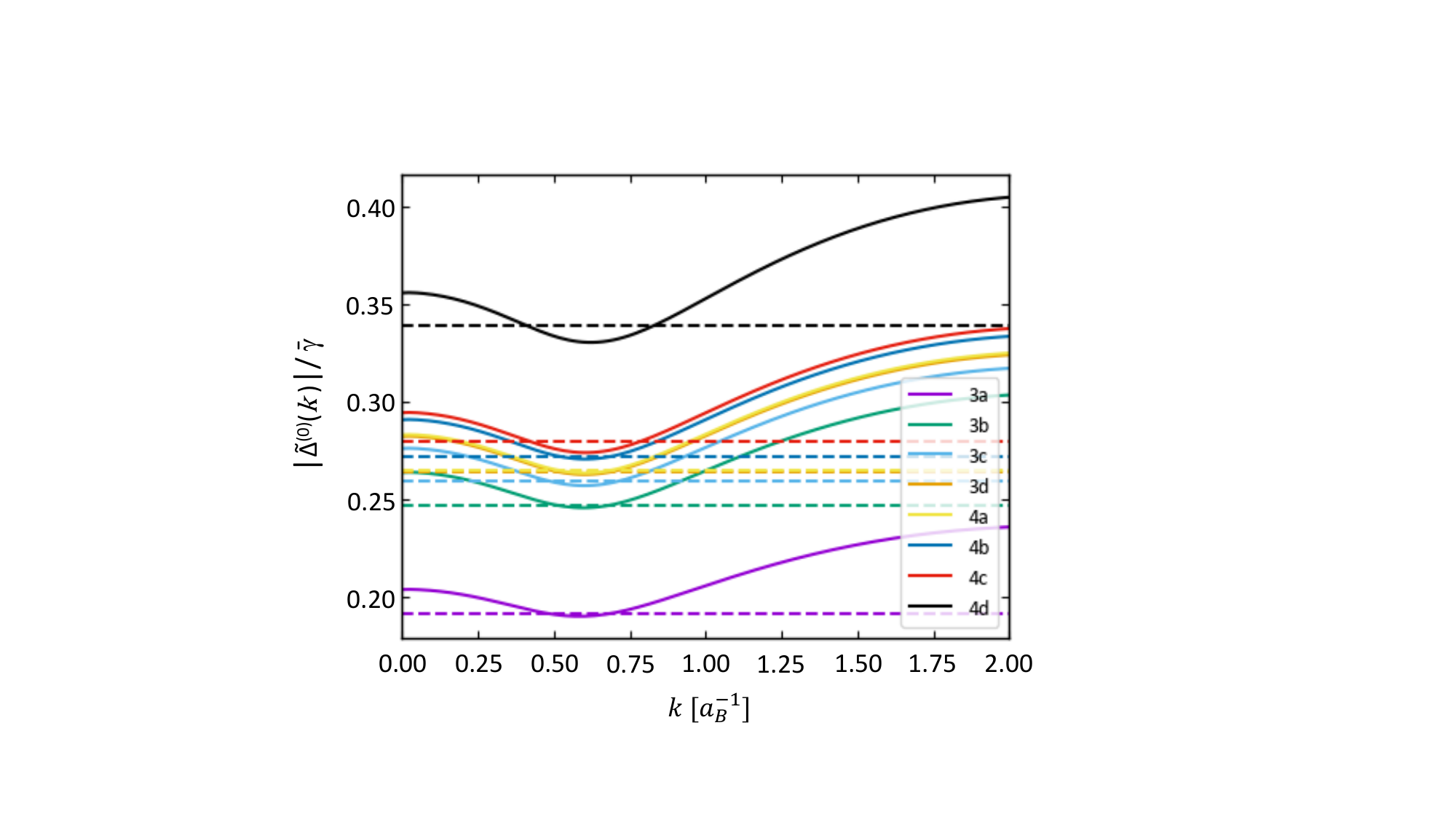}
\caption{Solid lines: Order parameter $| \tilde{\Delta}^{(0)} (k)|$ in the model with screened Coulomb interactions among the charged particles (defined by Eq. (\ref{Delta-0-def.equ})) plotted as a function of the electron momentum $k$, in units of inverse effective Bohr radius $a_B^{-1}$,for the eight cases shown in Figs. \ref{spectra-3.fig} and \ref{spectra-4.fig}.
Dashed lines: Values of the effective constant order parameter $| \Delta^{(0)}_{\textit{eff}} |$) used in the non-interacting electron-hole model to fit the eigenvalues of the interacting electron-hole model.
}
\label{Delta-vs-k.fig}
\end{figure}

\section{Interacting electrons and holes}
\label{THz-full-model.sec}

The Coulomb interaction not only models the charge carriers more realistically; it also provides an even clearer mathematical analogy to superconductivity than the other source of electron correlation, the coherent field, does alone.
To account for the presence of electrostatic interactions amongst and between electrons and holes requires, in general, that the full fluctuation equation \eqref{integral-equation.equ} be used.
The response matrix function $G_m (k,k^{\prime})$ cannot generally be assumed to be block diagonal in $k$, as is done in Eq. \eqref{THz-zero-V-G.equ}, because of the terms containing the order parameter fluctuation $\tilde{\Delta}^{(1)} (k,m,t)$ \eqref{Delta-1-def.equ} and the Hartree-Fock fluctuation $\Sigma^{(1)}_{HF} (k,m,t)$ \eqref{Sigma-1-def.equ}, which correlate the $p^{(1)}_{eh}$ and $f^{(1)}_e$ at different $k$'s.
One important effect caused by these $k$-mixing terms on the fluctuation spectrum is the possible formation of collective modes, where multiple $k$-states are excited simultaneously and which appear as discrete points in the complex plane. These have been studied in our previous work.\cite{binder-kwong.2021,spotnitz-etal.23}
In this paper, we examine order parameters smaller
than those necessary for the emergence of the THz collective ``T'' modes found in Ref. [\onlinecite{spotnitz-etal.23}].

To calculate the fluctuation continua for the exact response function $G_m (k,k^{\prime})$,
we numerically diagonalize the full $3 N_k \times 3 N_k$ matrix.
Here $N_k = 600$ is the number of $k$ grid points used.
(Otherwise, the parameters used for the
numerical calculations are the same as in Ref. [\onlinecite{binder-kwong.2021}], with effective Bohr radius $a_B = 14$ nm, electron temperature $T = 50$ K, and screening wavenumber $\kappa_0 = 9.0 \times 10^{-3} a_{B}^{-1}$.)
Using a weakly screened Coulomb interaction \cite{binder-kwong.2021,spotnitz-etal.23}, the eigenvalues are plotted as dots in Figs. \ref{spectra-3.fig} and \ref{spectra-4.fig}.
The EPs, as well-defined transition points, are very sensitive to the order parameter. To obtain the clear identification of the onset of an EP, as rendered in Figs. \ref{spectra-3.fig} and \ref{spectra-4.fig}, requires precisions on the order of 1 in $10^{5}$.

Remarkably, our numerical evidence indicates that the $k$-mixing terms % first-order perturbative quantities $\tilde{\Delta}^{(1)}$ and $\Sigma^{(1)}_{HF}$
	have little effect on the continuous branches of the fluctuation spectra.
Thus, for our purpose of investigating the parametric dependence of the eigenvalue continua, $\tilde{\Delta}^{(1)}$ and $\Sigma^{(1)}_{HF}$ can be very well approximated as $0$.
(The $k$-mixing terms significantly affect the eigenvectors, in that they are delta functions in $k$ in their absence, while in their presence they contain a range of $k$-values,\cite{spotnitz-etal.23}.
 This modification corroborates that the linear response is not simply weakly interacting.)
Under this approximation, the response matrix is block-diagonal in $k$ and is similar in structure to the matrix $M$ (Eq. \eqref{M-def.equ}) in the free electron-hole model:
\begin{equation}
	\label{M-int-def.equ}
	M_{int} (k) =
	\begin{pmatrix}
		\tilde{\xi}(k) - i \gamma_p & 0 & 2 \tilde{\Delta}^{(0)} (k) \\
		0 & - \tilde{\xi}(k) - i \gamma_p & - 2 \tilde{\Delta}^{(0) \ast} (k) \\
		\tilde{\Delta}^{(0) \ast} (k) & - \tilde{\Delta}^{(0)} (k) & - i \gamma_f
	\end{pmatrix}
\end{equation}
The fluctuation spectra calculated by diagonalizing the $3 \times 3$ matrix $M_{int}$ in Eq. \eqref{M-int-def.equ} for each $k$ are plotted as solid lines in Figs. \ref{spectra-3.fig} and \ref{spectra-4.fig}.
We see clearly that the dots, which represent the numerical eigenvalues in the continuous part of the spectrum, always overlap with the solid lines.
This invariance of the continuous spectrum under the action of the $k$-mixing terms is analogous to the invariance of the kinetic energy spectrum under the action of the scattering interaction Hamiltonian in quantum mechanics.\cite{taylor.1972}

Unlike in the noninteracting case, including Coulomb interactions generally means that the effective Rabi frequency, i.e. the order parameter, $\tilde{\Delta}^{(0)} (k)$ (defined in Eq. \eqref{Delta-0-def.equ}) is $k$-dependent.
However, let us consider a schematic contact potential, which is constant in $k$, up to a high-momentum cutoff $k_m$: $V (k,k') = V_{c} \, \theta(k_m - k) \theta(k_m - k')$, where $\theta$ is the unit step function.
Then the order parameter simplifies to a $k$-independent effective Rabi frequency $\tilde{\Delta}^{(0)}$. In this case, $M_{int}$ in Eq. \eqref{M-int-def.equ} simplifies to the $M$ of Eq. \eqref{M-def.equ}, with the only difference being the Rabi frequency $\Delta^{(0)}$ and electron-hole energy $\xi$ in \eqref{M-def.equ} are replaced in \eqref{M-int-def.equ} by the effective Rabi frequency $k$-constant $\tilde{\Delta}^{(0)}$ and the Hartree-Fock electron-hole energy $\tilde{\xi}$ (def. in Eq. \eqref{xi-def.equ}).
Thus, for a contact potential, the exact same continuous fluctuation spectra are obtained as in the free electron-hole model, and can be calculated and analyzed in the same way as in Section \ref{THz-linear-response.sec}, for the same choices of $\bar{\gamma}$, $\xi = \tilde{\xi}$, and $|\Delta| = |\tilde{\Delta}^{(0)}|$.

For a more realistic screened Coulomb interaction,\cite{binder-kwong.2021,spotnitz-etal.23} the eigenvalue curves are excellently approximated by diagonalization of Eq. \eqref{M-int-def.equ} with its $k$-dependent $\tilde{\Delta}^{(0)} (k)$, but the spectra are incomplete if approximated by Eq. \eqref{eigen.equ} with a $k$-independent effective Rabi frequency $\tilde{\Delta}^{(0)}$.
To show this, we have plotted the latter as dashed lines and (as stated previously) plotted the former with solid lines in Figs. \ref{spectra-3.fig} and \ref{spectra-4.fig}.
To pick the most appropriate $k$-constant $\tilde{\Delta}^{(0)}$ values, which we call the effective $\tilde{\Delta}^{(0)}$, $\Delta_{\textit{eff}}^{(0)}$, we simply varied $\Delta_{\textit{eff}}^{(0)}$ and re-plotted the dashed lines until they were visibly indistinguishable from the solid lines, or no visible improvement in the match between the two could be made.
We have plotted $\tilde{\Delta}^{(0)} (k)$ against $k$ in Fig. \ref{Delta-vs-k.fig}, with the $\Delta_{\textit{eff}}^{(0)}$ shown as dashed lines.
The identity of the $\Delta_{\textit{eff}}^{(0)}$ lines to portions of the $\tilde{\Delta}^{(0)} (k)$ lines in most of Figs. \ref{spectra-3.fig} and \ref{spectra-4.fig} shows that the $M_{int} (k)$ spectra contain the noninteracting spectra as a limiting case.
These dashed lines clearly reveal how the interacting case would reduce to the noninteracting case if the Coulomb interaction were to be turned off.

The increased complexity of Figures \ref{spectra-3.fig} and \ref{spectra-4.fig} relative to \ref{spectra-1.fig} and \ref{spectra-2.fig} suggests that these figures require additional explanation of the calculation and plotting methods.
All computations are done with a discretized $k$ grid, with bounds $0 \leq k \leq k_m$ and number of $k$ points $N_k$.
Due to the discretized $k$ grid, the continuous eigenvalues are calculated as a set of points distributed along the spectral continuum.
As the $N_k$ increases, these points shift and (usually -- see below for more discussion) become denser.
A discrete eigenvalue, in contrast, remains isolated and reaches convergence in position relatively fast with $k$ grid refinement. We have previously found these discrete eigenvalues to represent collective modes,\cite{binder-kwong.2021,spotnitz-etal.23} and do not study them in this paper.
Eigenvalue computations with $k$-mixing (i.e., for the complete response matrix) scale as
%$\mathcal{O}(N_k^2)$,
$\mathcal{O}(N_k^3)$,
while those without scale as $\mathcal{O}(N_k)$.
This performance improvement meant
that it was feasible to use a large enough $N_k$
such that the $k$-block-diagonal eigenvalues could be plotted as lines, 
 with effectively unbounded resolution,
while this resolution was relatively infeasible with $k$-mixing, and thus those results are plotted as points.
An additional limitation is that the $k$-mixing matrix was diagonalized numerically, while the eigenvalues of $M(k)$ and $M_{int}(k)$ were calculated by the cubic formula.
A matrix is not diagonalizable (defective) at an exceptional point,\cite{kato.95,ashida-etal.2020} while, in contrast, the multiplicity of roots poses no issue for solutions of a cubic equation.

Due to the complexity of the branches in the interacting case, it is clarifying to explicitly state how the continuous branch curves and their colors are chosen.
As each eigenvalue triplet is a solution of an independent $k$ equation, it is arbitrary how triplets from consecutive $k$-values are connected together to form branches, or whether this is even done at all.
We have decided to form the branches and assign the branch color by insisting on continuity of each branch in the $\Re \Omega$--$\Im \Omega$ plane.
The assignments of branches leaving EPs are arbitrary with respect to the branches entering them, so these were also chosen to ensure continuity, and with consistency across the different plots in mind.
We kept the color assignments consistent between the different $\Re \Omega$--$\Im \Omega$ plots as (the actual, or for the Coulomb case, effective) $|\Delta^{(0)}|$ value was varied by having the same colors assigned to the $k=0$ eigenvalues as they evolved smoothly across the plane. %for varying $|\Delta^{(0)}|$.
The $k=0$ points have been labeled with x's.
The branch assignment (colors) of the numerically-obtained eigenvalues (dots) were chosen to match that of the nearest $k$-block-diagonal fluctuation energies (nearest solid lines).

The Coulomb potential is $k$-dependent, which yields a $k$-dependent $\tilde{\Delta}^{(0)} (k)$, an example of which is shown in Fig. \ref{Delta-vs-k.fig}.
In the free electron-hole model, a constant $| \Delta |$ leads to a degeneracy between eigenvalues with $\xi \leq 0$ for $k < k_{\ell}$ and eigenvalues with $\xi \geq 0$ for $k > k_{\ell}$ (the ``low $k$'' and ``high $k$'' branches, respectively\cite{spotnitz-etal.23}) where $k_{\ell}$ is the momentum at which $\tilde{\xi}$ is zero, $\tilde{\xi} (k_{\ell}) = 0$.
This degeneracy is lifted by a $k$-dependent $\tilde{\Delta}^{(0)} (k)$.
The effect is visible in figures \ref{spectra-3.fig} and \ref{spectra-4.fig} as hook-like separations of branches that meet at the eigenvalue for $k_{\ell}$.

In Figs. \ref{spectra-3.fig} and \ref{spectra-4.fig}, the paths of the low-$k$ branches with increasing $k$ are denoted by brown arrows. The transition from low-$k$ to high-$k$ occurs at the $\tilde{\xi} (k_{\ell}) = 0$ points, which %, as stated previously,
are not explicitly marked in Figs. \ref{spectra-3.fig} and \ref{spectra-4.fig} but which occur at roughly the same positions as marked by small circles in Figs. \ref{spectra-1.fig} and \ref{spectra-2.fig}. Then, in Figs. \ref{spectra-3.fig} and \ref{spectra-4.fig}, the evolution of the high-$k$ branch is shown with purple arrows.

Increasing the number of branches makes the spectral features more complex than those in Figs. \ref{spectra-1.fig} and \ref{spectra-2.fig}.
As in Figs. \ref{spectra-1.fig} and \ref{spectra-2.fig}, the panels from (a) to (d) in Fig. \ref{spectra-3.fig} correspond to increasing overall strength of $\tilde{\Delta}^{(0)} (k)$, followed by the four panels of Fig. \ref{spectra-4.fig}.
The $k \geq k_{\ell}$ eigenvalues correspond very closely to the noninteracting case. This is confirmed from the match of the solid lines for these $k$-values with the entirety of the dashed lines, for most of figures \ref{spectra-3.fig} and \ref{spectra-4.fig}. In fact, this match is so precise, that the dashed lines are very difficult to distinguish from the solid lines in these cases.
The new behavior comes from the low-$k$ branch. Most notably, the lifted degeneracy of the low-$k$ branch creates a second set of 2nd- and 3rd-order EPs.

In Fig. \ref{spectra-3.fig}, panels (a)--(c), the red and green curves have been split into low- and high-$k$ branches. Their intersection now consists of two second-order EPs, one for each branch. A similar effect occurs with the green and blue curves in panels \ref{spectra-3.fig}(b) and \ref{spectra-3.fig}(c), so that these panels now have four second-order EPs. The two low-$k$ branch EPs move towards each other on the imaginary axis until they merge to form the first third-order EP in panel \ref{spectra-3.fig}(d). This first third-order EP marks a gapless-gapped transition for the low-$k$ branch, but the spectrum as a whole is still gapless.
As the pump density is marginally increased, the red and green branches separate from the imaginary axis, as shown in Fig. \ref{spectra-4.fig}(a). (To maintain continuity in this panel, it is necessary to have both the red and blue curves occupy a portion of the imaginary axis.)

Subsequently, as the pump density is further increased, the remaining pair of second-order EPs, which are in the high-$k$ branch, also move towards each other along the imaginary axis. They merge, as shown in Fig. \ref{spectra-4.fig}(b), to form the high-$k$ third-order EP. A marginal increase of the effective Rabi frequency $\left| \tilde{\Delta}^{(0)} \right|$, shown in Fig. \ref{spectra-4.fig}(c), prompts the red and green branches to fully separate from the imaginary axis, completing the transition from a gapless to a gapped spectrum. Further increases in the order parameter, as shown in Fig.\ \ref{spectra-4.fig}(d), increase the magnitude of the gap and move the location of minimal gap closer to the $\tilde{\xi} (k_{\ell}) = 0$ points.

Thus, the screened Coulomb interaction does not fundamentally modify the spectra. Its effect is to create a $k$-dependence in the order parameter $\left| \tilde{\Delta}_k^{(0)} \right|$. This variation lifts the low-$k$--high-$k$ degeneracy. The separated low-$k$ branch widens the spectra and introduces a second set of EPs, which also leads to a two-stage gapless-gapped transition with two third-order EPs, near to order parameters where in the noninteracting case there is only a single transition.

\section{Conclusion}

We have shown that there is in general a gapless regime in the single-particle (continuum) part of the THz fluctuation spectrum of a semiconductor microcavity laser. The gap refers to the separation of two branches of the spectral continuum from the imaginary axis. This gapless regime is an analog, in a driven-dissipative system, of the gapless regime in the single-particle excitation spectrum in some classes of superconductors.  The condition for the gapless regime considered here to exist is that the loss/relaxation rate $\gamma_f$ of the electron and hole occupation distribution is different from the dephasing rate $\gamma_p$ of the interband polarization. The conventional picture of the lasing order parameter $\Delta^{(0)} (\vb{k})= \Gamma_{e h} E^{(0)}_{\ell} + \Sigma_{\vb{k}^{\prime}} V_{\vb{k}-\vb{k}^{\prime}} p^{(0)}_{e h} (\vb{k}^{\prime})$ and the spectral gap having the same onset threshold is valid only in the limit $\gamma_f = \gamma_p$. In general, when $\gamma_f \neq \gamma_p$, the spectral gap opens at a finite value of $|\Delta^{(0)} (\vb{k})|$. As complex functions of $|\Delta^{(0)} (k)|$ and the electron-hole energy $\xi (k)$, the fluctuation eigenvalues show an interesting exceptional points structure, which we analyze in detail in both the photon laser limit and in the full model with Coulomb interactions among the charges. The gapless-gapped transition point is a third-order exceptional point.

The difference in dissipative rates, $\bar{\gamma} = \gamma_f - \gamma_p$, which controls the gapless regime, is a phenomenological parameter in this paper. These rates sum up the effects of many microscopic processes such as electron collisions, phonon emission/absorption. More insight into the physics of the gapless fluctuations may be gained by considering extensions of the current model that include microscopic descriptions of the dissipative processes. Other interesting directions for further studies include generalizing the theory to accommodate fluctuations with finite momentum (driven by obliquely incident probes) and analyzing analogous gapless spectra in the $m=0$ angular channel driven by optical-frequency interband probes.

We finally note that the physics discussed above is not necessarily restricted to our example of semiconductor lasers.
Future extensions of our work may include studies of optical quantum gases (e.g. \cite{ozturk-etal.2021}),
polymer lasers (e.g. \cite{samuel.2004}),
fiber-optic lasers (e.g. \cite{snitzer.1988,kieu-mansuripur.2008}), and other physical systems that can be described in terms of interacting or non-interacting two-level systems.

\begin{acknowledgments}

We gratefully acknowledge useful discussions with Hui Deng, University of Michigan,
financial support from the NSF under grant number DMR 1839570,
and the use of High Performance Computing (HPC) resources supported by the University of Arizona.

\end{acknowledgments}

\appendix

\section{Dynamical equations of the linear fluctuations of the laser}
\label{laser-fluctuation.sec}

In this appendix, we write down the linear response equations of the semiconductor laser under irradiation by a weak electromagnetic probe. Our general model equations to describe the laser dynamics are a set of semiconductor Bloch equations with phenomenological gain (pump) and loss (relaxation, dephasing, non-radiative decay) terms added on, and a probe is applied. The equations are expanded around a steady lasing state to first order in the probe. These equations are discussed in Refs. [\onlinecite{spotnitz-etal.21,spotnitz-etal.23}]. We reproduce the first order equations here.

The dynamical fields are the electron distribution $f_e (\vb{k}, t) \equiv \langle \hat{a}^{\dagger}_{e \vb{k}} (t) \hat{a}_{e \vb{k}} (t) \rangle$, the interband polarization $p_{e h} (\vb{k}, t) \equiv \langle \hat{a}_{h (-\vb{k})} (t) \hat{a}_{e \vb{k}} (t) \rangle$, and the uniform cavity photon field $E_{\ell} (t)$. The hole distribution is assumed to be the same as the electron distribution. We assume that the laser steady state is set in a circularly polarized state and suppress the spin and polarization direction labels. The probe is a normally incident plane wave.
Each field is written as a sum of two terms:
\begin{equation}
\Psi(\vb{k}, t) = \Psi^{(0)} (\vb{k}) + \Psi^{(1)} (\vb{k}, t) \quad , \quad \Psi = f_e , p_{e h} \quad , \quad E_{\ell} (t) = E^{(0)}_{\ell} + E^{(1)}_{\ell} (t)
\end{equation}
where the $(0)$ superscript denotes the lasing steady state and the $(1)$ superscript denotes the fluctuation response of the laser to the probe.
The $\vb{k}$ dependent fluctuation fields are expanded in angular harmonics ($\vb{k} = (k , \theta_k)$):
\begin{equation}
\Psi^{(1)} (k,\theta_k , t) = \sum_{m \in \mathbb{Z}} \Psi^{(1)} (k,m,t) e^{im\theta_k}
\quad , \quad
\Psi^{(1)}(k,m,t) = \frac{1}{2\pi}\int_{0}^{2\pi} d\theta_k \Psi^{(1)} (k,\theta_k,t) e^{-im\theta_k}
\label{modeexpans.equ}
\end{equation}
where $\Psi^{(1)}$ stands for $p^{(1)}_{e h}$, $f^{(1)}_{e}$. Being independent of $\vb{k}$, $E_{\ell}^{(1)}$ is circularly symmetric.
The equations for individual $m$ components of the charge carrier fields are
%\begin{widetext}
\begin{eqnarray}
i\hbar \frac{\partial }{\partial t} p^{(1)}_{e h}(k,m,t) &=&
\left( \tilde{\xi} (k) - i \gamma_p \right) p^{(1)}_{e h}(k,m,t)
+ 2 p^{(0)}_{e h}(k) \Sigma^{(1)}_{HF} (k,m,t) \nonumber \\
&& - \left[ 1-2f^{(0)}_{e}(k)\right] \tilde{\Delta}^{(1)} (k,m,t)
+ 2 \tilde{\Delta}^{(0)} (k) f^{(1)}_{e}(k,m,t) \nonumber \\
&& + S(k,m,t) \label{delta-P-dot-ang.equ}
\end{eqnarray}
\begin{eqnarray}
-i\hbar \frac{\partial }{\partial t}p^{(1) \ast}_{e h}(k,-m,t) &=&
\left( \tilde{\xi} (k) + i \gamma_p \right) p^{(1) \ast}_{e h}(k,-m,t)
- 2 p^{(0) \ast}_{e h}(k) \Sigma^{(1) \ast}_{HF} (k,-m,t) \nonumber \\
&& - \left[ 1-2f^{(0)}_{e}(k)\right] \tilde{\Delta}^{(1)\ast} (k,-m,t)
+ 2 \tilde{\Delta}^{(0) \ast} (k) f^{(1) \ast}_{e}(k,-m,t) \nonumber \\
&& + S^{\ast}(k,-m,t) \label{delta-P-star-dot-ang.equ}
\end{eqnarray}
\begin{eqnarray}
i \hbar \frac{\partial }{\partial t}f^{(1)}_{e}(k,m,t) &=&
\tilde{\Delta}^{(0) \ast} (k) p^{(1)}_{e h}(k,m,t) +
p^{(0)}_{e h}(k) \tilde{\Delta}^{(1)\ast} (k,-m,t) \nonumber \\
&& - \tilde{\Delta}^{(0)} (k) p^{(1) \ast}_{e h}(k,-m,t) -
p^{(0) \ast}_{e h}(k) \tilde{\Delta}^{(1)} (k,m,t) \nonumber \\
&&-i \gamma_{f}  f^{(1)}_{e}(k,m,t) \label{delta-f-dot-ang.equ}
\end{eqnarray}
and the equations for the cavity laser field are
\begin{equation}
i\hbar \frac{\partial }{\partial t}E_{\ell}^{(1)} (t) =\left( \hbar \omega _{cav}-\hbar
\omega _{\ell}-i\gamma _{cav}\right) E_{\ell}^{(1)} (t) - N_{QW} \int_{0}^{\infty} \frac{k \mathrm{d}k}{2\pi} \Gamma^{\ast}_{e h} (k)p^{(1)}_{e h}(k,0,t)
\label{delta-E-dot-ang.equ}
\end{equation}
\begin{equation}
-i\hbar \frac{\partial }{\partial t}E_{\ell}^{(1)\ast} (t) =\left( \hbar \omega_{cav}-\hbar \omega _{\ell}+i\gamma _{cav}\right) E_{\ell}^{(1)\ast} (t) - N_{QW} \int_{0}^{\infty} \frac{k \mathrm{d}k}{2\pi} \Gamma^{}_{e h} (k)p^{(1) \ast}_{e h}(k,0,t) \label{delta-E-star-dot-ang.equ}
\end{equation}
The symbols in these equations are defined as follows. $\tilde{\xi} (k)$ is the sum of the electron and hole single-particle energies and is given by
\begin{equation}
\tilde{\xi} (k) = \frac {\hbar^2 k^2} {m} + E_g + 2 \Sigma^{(0)}_{HF} (k) - \hbar \omega_{\ell}
\label{xi-def.equ}
\end{equation}
where $m$ is the common effective mass of the electron and the hole, $E_g$ is the band gap, and $\omega_{\ell}$ is the lasing frequency.
$\tilde{\Delta}^{(0)} (k)$ and $\tilde{\Delta}^{(1)} (k,m,t)$ are the steady-state and fluctuation components of the order parameter, and $\Sigma^{(0)}_{HF} (k)$ and $\Sigma^{(1)}_{HF} (k,m,t)$ are the components of the Hartree-Fock energy. They are given by
\begin{align}
\tilde{\Delta}^{(0)} (k) &= \Gamma_{e h} (k)E_{\ell}^{(0)} + \int_{0}^{\infty} \frac{k^{\prime} \mathrm{d} k^{\prime}}{2\pi} V_{k,k^{\prime}}^{0} p^{(0)}_{e h}(k^{\prime})
\label{Delta-0-def.equ} \\
\tilde{\Delta}^{(1)} (k,m,t) &= \Gamma_{e h} (k)E_{\ell}^{(1)} (t) \delta_{0,m} +\int_{0}^{\infty} \frac{k^{\prime} \mathrm{d} k^{\prime}}{2\pi} V_{k,k^{\prime}}^{m} p^{(1)}_{e h}(k^{\prime},m,t)
\label{Delta-1-def.equ} \\
\Sigma^{(0)}_{HF} (k) &= - \int_{0}^{\infty} \frac{k^{\prime} \mathrm{d}k^{\prime}}{2\pi} V_{k,k^{\prime}}^{0} f^{(0)}_{e} (k^{\prime}) \label{Sigma-0-def.equ} \\
\Sigma^{(1)}_{HF} (k,m,t) &= - \int_{0}^{\infty} \frac{k^{\prime} \mathrm{d} k^{\prime}}{2\pi} V_{k,k^{\prime}}^{m}  f^{(1)}_{e}(k^{\prime},m,t) \label{Sigma-1-def.equ}
\end{align}
$\Gamma_{e h} (k)$ is the coefficient of the coupling that transforms a cavity photon into an electron-hole pair. $V^{m}_{k k^{\prime}}$ is the component with angular momentum $m$ of the electron-electron interaction with incoming relative momentum $\vb{k}^{\prime}$ and outgoing relative momentum $\vb{k}$. A statically screened Coulomb potential is usually chosen to be the interaction. For our definition of $V^{m}_{k k^{\prime}}$ here, we only need to assume that the interaction depends on $\vb{k} - \vb{k}^{\prime}$ instead of $\vb{k}$ and $\vb{k}^{\prime}$ separately, which implies a dependence on only the three scalars $k, k^{\prime}, \theta$, where $\theta$ is the angle between $\vb{k}$ and $\vb{k}^{\prime}$. If the interaction is denoted by $V (k , k^{\prime} , \theta)$, the angular component is given by
\begin{equation}
V^{m}_{k k^{\prime}} = \frac {1} {2 \pi} \int_{0}^{2 \pi} \mathrm{d} \theta e^{- i m \theta} V (k , k^{\prime} , \theta)
\end{equation}
$\gamma_p$, $\gamma_f$, and $\gamma_{cav}$ are the loss/relaxation rates of the interband polarization, the occupation, and the cavity field respectively. $\omega_{cav}$ is the resonance frequency of the cavity mode, and $N_{QW}$ is the number of quantum wells in the cavity. $S(k,m,t)$ represents the $m$ angular component of the external probe field. For an optical-frequency probe, which effects interband transitions, the $m=0$ component is dominant, and the other components are negligibly small: $S(k,m,t) \sim \delta_{m 0}$. For a THz probe, which drives intraband motion, $S(k,m,t)$ is nonzero only for $m = \pm 1$.
More detailed discussion of these equations can be found in Refs. [\onlinecite{spotnitz-etal.21,spotnitz-etal.23}].

Eqs. (\ref{delta-P-dot-ang.equ})-(\ref{delta-E-star-dot-ang.equ}) can be rewritten more compactly in the form of a matrix integral equation
\begin{equation}
i \hbar \frac {\partial} {\partial t} \vec{x} (k,m,t) = \int_{0}^{\infty} \mathrm{d} k^{\prime} G_m (k , k^{\prime}) \vec{x} (k^{\prime},m,t) + \vec{S} (k,m,t)
\label{integral-equation.equ}
\end{equation}
where $\vec{x} (k,m,t)$ is the five-dimensional column vector
\begin{equation}
\vec{x} (k,m,t) =
\begin{pmatrix}
 p^{(1)}_{e h} (k,m,t) \\ p^{(1) \ast}_{e h} (k,-m,t) \\ f^{(1)}_{e} (k,m,t) \\
 E^{(1)}_{\ell} (t) \delta_{m 0} \\ E^{(1) \ast}_{\ell} (t) \delta_{m 0}
\end{pmatrix}
\end{equation}
and $\vec{S}(k,m,t)$ is the column vector containing the probe terms. The 'response function' $G_m (k,k^{\prime})$ is a $5 \times 5$ matrix with each element being a function of $k$ and $k^{\prime}$. Its eigenmodes are the fluctuation modes of angular momentum $m$. Its spectrum consists of one or more continuous curve(s) in the complex plane and possibly also a set of discrete points (collective modes).\\

\noindent
\textit{The photon laser limit}\\

The matter components of the fluctuations, $p^{(1)}_{e h} (k,m,t)$ and $f^{(1)}_{e} (k,m,t)$, at different $k$ values are in general correlated. In Eqs. (\ref{delta-P-dot-ang.equ})-(\ref{delta-f-dot-ang.equ}) this correlation lies in terms carrying the order parameter fluctuation $\tilde{\Delta}^{(1)} (k,m,t)$ (defined in Eq. (\ref{Delta-1-def.equ})) and density fluctuation $\Sigma^{(1)}_{HF} (k,m,t)$ (defined in Eq. (\ref{Sigma-1-def.equ})). In cases where these terms vanish, decoupling in $k$ space occurs and simplifies the mode analysis significantly. In the photon laser limit, where the interaction effects can be neglected, $\Sigma^{(1)}_{HF} (k,m,t) = 0$, and the order parameter fluctuation is reduced to the coupling to the photon fluctuation: $\tilde{\Delta}^{(1)} (k,m,t) = \Gamma_{e h} (k)E_{\ell}^{(1)} (t) \delta_{0,m}$, which is nonzero only in the $m=0$ sector. So, specifically, the $k$ correlation is present for optical probe perturbations ($m = 0$) of a photon laser but absent for THz probe perturbations $m=\pm 1$.

In the matrix formulation Eq. (\ref{integral-equation.equ}), the response function $G_m (k, k^{\prime})$ becomes block-diagonal in $k$ for the THz-probed photon laser. Due to anglar momentum mismatch, the photon fluctuation fields $E^{(1)}_{\ell} (t)$ and $E^{(1) \ast}_{\ell} (t)$ vanish (there is no source $S$ for $p^{(1)}_{e h} (k,0,t)$ in Eqs. (\ref{delta-P-dot-ang.equ})--(\ref{delta-f-dot-ang.equ}), so the source term $p^{(1)}_{e h} (k,0,t) = 0$ in Eqs. (\ref{delta-E-dot-ang.equ}) and (\ref{delta-E-star-dot-ang.equ}) and $E^{(1)}_{\ell} (t)$ and $E^{(1) \ast}_{\ell} (t)$ are removed from the vector $\vec{x} (k,1,t)$). Setting $V^{m}_{k k^{\prime}} = 0$, we obtain
\begin{equation}
G_m (k, k^{\prime}) = \delta(k-k^{\prime}) M (k) \quad , \quad
M (k) =
\begin{pmatrix}
\xi(k) - i \gamma_p & 0 & 2 \Delta \\
0 & - \xi(k) - i \gamma_p & -2 \Delta^{\ast} \\
\Delta^{\ast} & - \Delta & - i \gamma_f
\end{pmatrix}
\label{THz-zero-V-G.equ}
\end{equation}
where
\begin{equation}
\xi (k) = \frac {\hbar^2 k^2} {m} + E_g - \hbar \omega_{\ell} \quad , \quad \Delta = \tilde{\Delta}^{(0)} ( V^{m}_{k,k^{\prime}} = 0) = \Gamma_{eh} E^{(0)}_{\ell}
\end{equation}
We assume $\Gamma_{eh}$ to be a constant so that the steady state order parameter is independent of $k$. Eq. (\ref{integral-equation.equ}) becomes block-decoupled, with the equations for each $k$ being
\begin{equation}
i \hbar \frac {\partial} {\partial t} \vec{x} (k,1,t) = M (k) \vec{x} (k,1,t) + \vec{S} (k,1,t)
\end{equation}
\begin{equation}
\vec{x} (k,1,t) =
\begin{pmatrix}
 p^{(1)}_{e h} (k,1,t) \\ p^{(1) \ast}_{e h} (k,-1,t) \\ f^{(1)}_{e} (k,1,t)
\end{pmatrix}
\quad , \quad
\vec{S} (k,1,t) =
\begin{pmatrix}
S (k,1,t) \\ - S^{\ast} (k,-1,t) \\ 0
\end{pmatrix}
\end{equation}

\section{Properties of the eigenvalues}
\label{derivation.sec}

In this Appendix, we derive some properties of the fluctuation eigenvalue spectrum of the non-interacting electron-hole model (Section \ref{free-eh.sec}), in which the order parameter $\Delta$ is independent of $k$.
%We also indicate which properties hold also when $\Delta$ depends on $k$, which is the case when a $k$-dependent interaction, such as the Coulomb force, acts between the charges (Section \ref{THz-full-model.sec}).

In the non-interacting model, the eigenvalue equation (Eq. (\ref{eigen.equ}))
\begin{equation}
\label{eigen-app.equ}
( \bar{\lambda}^2 - \xi^2 )( \bar{\lambda} + i \bar{\gamma} ) - 4
|\Delta|^2 \bar{\lambda} = 0
\end{equation}
is a cubic equation with complex coefficients. Defining the real and imaginary parts of the solution
\begin{equation}
\bar{\lambda} = a + i b \quad , \quad a,b \in \mathbb{R}
\end{equation}
we break the eigenvalue equation into its real and imaginary parts
\begin{align}
a^3 - (3 b^2 + 2 \bar{\gamma} b + \xi^2 + 4 | \Delta |^2) a &= 0 \label{eigen-real.equ} \\
b^3 + \bar{\gamma} b^2 + ( \xi^2 + 4 | \Delta |^2 - 3 a^2 ) b +
\bar{\gamma} ( \xi^2 - a^2 ) &= 0 \label{eigen-imag.equ}
\end{align}
From Eq. (\ref{eigen-real.equ}) we express $a$ in terms of other quantities as
\begin{equation}\label{a.equ}
a = 0 \quad {\rm or} \quad a^2 = 3 b^2 + 2 \bar{\gamma} b + \xi^2 + 4 | \Delta |^2
\end{equation}
Substituting these two expressions into Eq. (\ref{eigen-imag.equ}) gives the following two sets of equations
\begin{align}
b^3 + \bar{\gamma} b^2 + ( \xi^2 + 4 | \Delta |^2) b +
\bar{\gamma} \xi^2 &= 0 \quad ; \quad a=0 \label{eigen-1.equ} \\
b^3 + \bar{\gamma} b^2 + \left[ \frac {\xi^2} {4} + | \Delta |^2 + \frac {\bar{\gamma}^2} {4} \right] b +
\frac {| \Delta |^2} {2} \bar{\gamma} &= 0 \quad ; \quad a^2 = 3 b^2 + 2 \bar{\gamma} b +
\xi^2 + 4 | \Delta |^2 \label{eigen-2.equ}
\end{align}
The equations for $b$ in both sets are cubic equations with real coefficients. Explicit solutions for this class of equations exist (e.g. Ref. [\onlinecite{abramowitz-stegun.72}]), which we use in the numerical calculations. Generally,
Eqs. (\ref{eigen-1.equ}) and (\ref{eigen-2.equ}) together have six solutions. The constraint that both
$b$ ($= {\rm Im} \bar{\lambda}$) and $a$ ($={\rm Re} \bar{\lambda} $) must be real numbers picks, out of the six, the three solutions that are also solutions of Eq. (\ref{eigen.equ}). The condition that $a$ is real implies that the right hand side of the second equation in Eq. (\ref{eigen-2.equ}) must be positive or equal to zero. A cubic equation with real coefficients has at least one real solution, and the other two solutions are either real or complex conjugates of each other \cite{abramowitz-stegun.72}. These considerations lead to the following three groups of allowed solutions:
(1) Eq. (\ref{eigen-1.equ}) yields three real solutions for $b$, which together with $a = 0$, give the three
eigenvalues in Eq. (\ref{eigen.equ}). Eq. (\ref{eigen-2.equ}) does not contribute any allowed solution. (2) Eq. (\ref{eigen-1.equ}) has one real-$b$ solution, and Eq. (\ref{eigen-2.equ}) has one solution with real $b$ and $a^2 > 0$. The latter gives the pair of eigenvalues off the imaginary axis. (3) At some particular values of $| \Delta |$ , $\xi$, Eq. (\ref{eigen-2.equ}) has one or more solutions with real $b$ and $a^2 = 0$. These solutions coincide with solutions of Eq. (\ref{eigen-1.equ}).

As functions of a constant ($k$-independent) $|\Delta|$, $\xi$, and $\bar{\gamma}$, the eigenvalues of Eq. (\ref{eigen-app.equ}) have the following properties:\\

\noindent
(1) If $\bar{\gamma} > 0$, the eigenvalues are confined to
${\rm Im} \bar{\lambda} \equiv b \in [ - \bar{\gamma} , 0 ]$ on the imaginary axis and
${\rm Im} \bar{\lambda} \in [ - \frac {\bar{\gamma}} {2} , 0 ]$ off the imaginary axis. The proof goes as follows.\\

\noindent
\textit{Claim}: sign($b$) = - sign($\bar{\gamma}$).\\
\textit{Proof}: Eq. (\ref{eigen-1.equ}) can be written as
\begin{equation}
(b^2 + \xi^2 + 4 | \Delta |^2) b + \bar{\gamma} (b^2 + \xi^2) = 0
\end{equation}
which gives
\begin{equation}
\frac {b} {\bar{\gamma}} = - \frac {b^2 + \xi^2}
{b^2 + \xi^2 + 4 | \Delta |^2}
\end{equation}
Since the right hand side is negative, we have sign($b$) = - sign($\bar{\gamma}$).

Eq. (\ref{eigen-2.equ}) can be written as
\begin{equation}
b \left[ b^2 + \frac {\xi^2} {4} + | \Delta |^2 + \frac {\bar{\gamma}^2} {4} \right] + \bar{\gamma}
\left[ b^2 + \frac {| \Delta |^2} {2} \right] = 0
\end{equation}
and the same argument as above applies. \\
\textit{Claim}: If $\bar{\gamma} > 0$ and $a = 0$, then $b \ge -\bar{\gamma}$.\\
\textit{Proof}: Write Eq. (\ref{eigen-1.equ}) as
\begin{equation}
(b^2 + \xi^2) (b + \bar{\gamma}) + 4 | \Delta |^2 b = 0
\end{equation}
If $b < -\bar{\gamma}$, both terms on the right hand side are negative, and the equation cannot be
satisfied. So $b \ge -\bar{\gamma}$.\\
\textit{Claim}: If $\bar{\gamma} > 0$ and $a^2 > 0$, then $b \ge - \frac {\bar{\gamma}} {2}$.\\
\textit{Proof}: Write Eq. (\ref{eigen-2.equ}) as
\begin{equation}\label{eigen-2a.equ}
b^2 (b + \bar{\gamma}) + \frac {1} {4} ( \xi^2 + \bar{\gamma}^2) b + \left[ b + \frac {\bar{\gamma}}
{2} \right] | \Delta |^2 = 0
\end{equation}
Suppose $b < - \frac {\bar{\gamma}} {2}$. The third term on the left hand side is negative. Consider
the sum of the first two terms and remove the common factor $b$:
\begin{equation}
b (b + \bar{\gamma}) + \frac {1} {4} ( \xi^2 + \bar{\gamma}^2) \end{equation}
As a function of $b$, the first term is a parabola with a minimum at $b = - \frac {\bar{\gamma}} {2}$.
The corresponding minimum value of the parabola is $\min{[b (b + \bar{\gamma})]} = - \frac {\bar{\gamma}^2}{4}$.
This gives
\begin{equation}
b (b + \bar{\gamma}) + \frac {1} {4} ( \xi^2 + \bar{\gamma}^2) \ge \frac {\xi^2} {4} \ge 0
\end{equation}
Since $b$ is negative by assumption, the sum of the first two terms in Eq. (\ref{eigen-2a.equ}), which is the
above expression multiplied by $b$, is less than or equal to 0. So if $b < - \frac {\bar{\gamma}} {2}$,
the left hand side of Eq. (\ref{eigen-2a.equ}) is negative and the equation cannot be satisfied.
So the claim is proved.

The three claims, taken together, prove the statement of the property.\\

\noindent
(2) Eigenvalues at $\xi = 0$: $\bar{\lambda} = 0$ is always an eigenvalue. The other two eigenvalues are purely imaginary when $|\Delta| \le \frac {\bar{\gamma}} {4}$. When $|\Delta| > \frac {\bar{\gamma}} {4}$, they have non-zero real parts and their imaginary parts are ${\rm Im} \bar{\lambda} = - \frac {\bar{\gamma}} {2}$.\\

\noindent
\textit{Proof}: When $\xi=0$, Eq. (\ref{eigen-1.equ}) becomes
\begin{equation}
( b^2 + \bar{\gamma} b  + 4 | \Delta |^2 ) b = 0 \quad , \quad a=0
\end{equation}
from which it is clear that $\bar{\lambda} = 0$ is a solution for all values of $| \Delta |$. The other solutions are
\begin{equation}
b = - \frac {\bar{\gamma}} {2} \pm \sqrt{\left( \frac {\bar{\gamma}} {2} \right)^2 - 4 | \Delta |^2}
\end{equation}
These solutions satisfy the real-$b$ constraint when $| \Delta | \in [0 , \frac {\bar{\gamma}} {4} ]$.
At $| \Delta | = 0$, the two solutions are $b = 0 , - \bar{\gamma}$. As $| \Delta |$ increases, they
converge and coincide at the value of $b = - \frac {\bar{\gamma}} {2}$
when $| \Delta | = \frac {\bar{\gamma}} {4}$.
For $| \Delta | > \frac {\bar{\gamma}} {4}$, the allowed solutions other than $\bar{\lambda} = 0$ are
given by Eq. (\ref{eigen-2.equ}) which reduces at $\xi = 0$ to
\begin{equation}
b^3 + \bar{\gamma} b^2 + \left[ | \Delta |^2 + \frac {\bar{\gamma}^2} {4} \right] b +
\frac {| \Delta |^2} {2} \bar{\gamma} = 0 \quad ; \quad a^2 = 3 b^2 + 2 \bar{\gamma} b +
4 | \Delta |^2 \label{eigen-2b.equ}
\end{equation}
It can be verified by direct substitution that
\begin{equation}
b = - \frac {\bar{\gamma}} {2} \quad , \quad a^2 = - \frac {\bar{\gamma}^2} {4} + 4 | \Delta |^2
\label{xi0abDelgquart.equ}
\end{equation}
is a solution of Eq. (\ref{eigen-2b.equ}). Note that the $b = - \frac {\bar{\gamma}} {2}$ part of the solution is valid for all values of $| \Delta |$. The requirement of $a^2 \ge 0$ implies that these solutions are allowed
only for $| \Delta | > \frac {\bar{\gamma}} {4}$.\\

\noindent
(3) At $\xi \rightarrow \infty$, one eigenvalue tends to $-i \bar{\gamma}$ and the other two diverge as $\pm \xi$.\\

\noindent
\textit{Proof}: For eigenvalues on the imaginary axis, as $\xi \rightarrow \infty$, the left hand side of
Eq. (\ref{eigen-1.equ}) is dominated by the terms containing $\xi^2$ (the other terms being
bounded because $b \in [- \bar{\gamma} , 0]$). The equation then reduces to
\begin{equation}
\xi^2 (b + \bar{\gamma}) = 0 \quad \Rightarrow \quad b \rightarrow - \bar{\gamma}
\end{equation}
For eigenvalues off the imaginary axis, as $\xi \rightarrow \infty$, Eq. (\ref{eigen-2.equ}) reduces to
\begin{equation}
\frac {\xi^2} {4} b = 0 \quad \Rightarrow \quad b \rightarrow 0 \quad , \quad a \rightarrow \pm \xi
\end{equation}\\

\noindent
(4) $\bar{\lambda} = - i \frac {\bar{\gamma}} {3}$ is a triply degenerate solution at
$| \Delta | = \sqrt{\frac {2} {27}} \bar{\gamma}$ and $\xi = \sqrt{\frac {1} {27}} \bar{\gamma}$.\\

\noindent
\textit{Proof}: Let $b = b_0 , a = 0$ be a candidate triply degenerate solution. The cubic equation then has the following form
\begin{equation}
( b - b_0 )^3 = b^3 - 3 b_0 b^2 + 3 b_0^2 b - b_0^3 = 0
\end{equation}
Identifying the coefficients of this equation with those of Eq. (\ref{eigen-1.equ}) gives the conditions that
$b_0$, $|\Delta|$, and $\xi$ must satisfy:
\begin{align}
- 3 b_0 &= \bar{\gamma} \\
3 b_0^2 &= \xi^2 + 4 | \Delta |^2 \\
- b_0^3 &= \bar{\gamma} \xi^2
\end{align}
The first equation gives $b_0 = - \frac {\bar{\gamma}} {3}$, and with this value substituted into the other two
equations, they give the values of $\xi = \sqrt{\frac {1} {27}} \bar{\gamma}$ and $| \Delta | = \sqrt{\frac {2} {27}} \bar{\gamma}$. These values, together with $a=0$, also satisfy Eq. (\ref{eigen-2.equ}).\\

\noindent
(5) When $| \Delta | \ge \frac {\bar{\gamma}} {\sqrt{8}}$, the minimum of
${\rm Re} \bar{\lambda}$ over the set of $\xi$ values is at $\xi = 0$. (The value of
${\rm Re} \bar{\lambda}$ at $\xi = 0$ equals the gap.)\\

\noindent
\textit{Proof}: We look for the value of $| \Delta |$ where the slope of $a^2$ as a function of $\xi^2$ changes
from negative to positive, i.e.
\begin{equation}
\left. \frac {\partial a^2} {\partial \xi^2} \right|_{\xi = 0} = 0
\end{equation}
The expression for $a^2$ in Eq. (\ref{eigen-2.equ}) is
\begin{equation}
a^2 = 3 b^2 + 2 \bar{\gamma} b +
\xi^2 + 4 | \Delta |^2
\end{equation}
In this expression $b$ is also a function of $\xi$ while $\bar{\gamma}$ and $| \Delta |$ are parameters.
The derivative is
\begin{equation}\label{a-xi-deriv.equ}
\frac {\partial a^2} {\partial \xi^2} = (6 b + 2 \bar{\gamma}) \frac {\partial b} {\partial \xi^2} + 1
\end{equation}
The $\xi^2$-derivative of $b$ is obtained by taking the $\xi^2$-derivative of the
cubic equation in Eq. (\ref{eigen-2.equ}):
\begin{equation}
\left[ 3 b^2 + 2 \bar{\gamma} b + \frac{1} {4} (\xi^2 + \bar{\gamma}^2) + | \Delta |^2 \right]
\frac {\partial b} {\partial \xi^2} + \frac {b} {4} = 0
\end{equation}
giving
\begin{equation}\label{b-xi-deriv.equ}
\frac {\partial b} {\partial \xi^2} = - \frac {b} {4}
\left[ 3 b^2 + 2 \bar{\gamma} b + \frac{1} {4} (\xi^2 + \bar{\gamma}^2) + | \Delta |^2 \right]^{-1}
\end{equation}
From Eq. \eqref{xi0abDelgquart.equ}, at $\xi = 0$, $b = - \frac {\bar{\gamma}} {2}$. Substituting these values into Eq. (\ref{b-xi-deriv.equ}) gives
\begin{equation}
\left. \frac {\partial b} {\partial \xi^2} \right|_{\xi = 0} = \frac {\bar{\gamma}} {8 | \Delta |^2}
\end{equation}
With these values, Eq. (\ref{a-xi-deriv.equ}) becomes
\begin{equation}
\left. \frac {\partial a^2} {\partial \xi^2} \right|_{\xi = 0} = 1 - \frac {\bar{\gamma}^2} {8 | \Delta |^2}
\end{equation}
Setting $\left. \frac {\partial a^2} {\partial \xi^2} \right|_{\xi = 0} = 0$ gives
\begin{equation}
| \Delta |^2 = \frac {\bar{\gamma}^2} {8}
\end{equation}\\
\noindent
Properties (2), (4), and (5) give critical values of $| \Delta |$  (= $\frac {\bar{\gamma}} {4}$,
$\sqrt{\frac {2} {27}} \bar{\gamma}$, and $\sqrt{\frac {1} {8}} \bar{\gamma}$) where the behavior of
the eigenvalue set changes. Together with numerical results, they produce the qualitative picture of the
eigenvalues' dependence on $| \Delta |$ summarized in Section \ref{free-eh.sec}.

\bibliographystyle{prsty_noetal_noabbr}

%Nai
%\bibliography{gap,allref,TSSO-Gapless}

%Rolf
%\bibliography{../../bib/allref,gap,TSSO-gapless}%

\end{document}